%% file: JASA-template.tex
\documentclass[11pt]{article}
\usepackage{amsmath}
\usepackage{graphicx,psfrag,epsf}
\usepackage{enumerate}
\usepackage{natbib}
\usepackage{url} 
\usepackage[margin=1in]{geometry}
\usepackage{amsfonts}
\usepackage{algorithm}
\usepackage{algpseudocode}
\usepackage{dsfont}
\usepackage{color}
\usepackage[T1]{fontenc}
\usepackage[font=footnotesize]{caption}

\newcommand{\blind}{0}

\newcommand{\R}{{\mathds R}}
\newcommand{\x}{{x}}
\newcommand{\X}{{X}}

\newcommand{\Norm}{\mathcal{N}}
\newcommand{\iv}{\mathbb{I}} 
\newcommand{\E}{\mathds{E}} 
\renewcommand{\P}{\mathds{P}} 
\newcommand{\PM}{\textsc{pm}$_{2.5}$ }
\newcommand{\PMns}{\textsc{pm}$_{2.5}$}

\newcommand\blfootnote[1]{%
  \begingroup
  \renewcommand\thefootnote{}\footnote{#1}%
  \addtocounter{footnote}{-1}%
  \endgroup
}

\usepackage[table]{xcolor}  
\definecolor{lightGray}{HTML}{DEDEDE}
\definecolor{Gray2}{HTML}{F7F7F7}

\begin{document}

\def\spacingset#1{\renewcommand{\baselinestretch}%
{#1}\small\normalsize} \spacingset{1}


\if1\blind
{\title{\bf Confounder-Dependent Bayesian Mixture Model: Characterizing Heterogeneity of  Causal Effects in Air Pollution Epidemiology}
 \author{Dafne Zorzetto \thanks{
 The authors gratefully acknowledge \textit{please remember to list all relevant funding sources in the unblinded version}}\hspace{.2cm}\\
 Department of Statistics, University of Padova, Italy.\\
 Falco J. Bargagli-Stoffi \\
 Department of Biostatistics, Harvard T.H. Chan School of Public Health, Massachusetts, USA \\
 Antonio Canale \\
 Department of Statistics, University of Padova, Italy.\\
 and \\
 Francesca Dominici \\
 Department of Biostatistics, Harvard T.H. Chan School of Public Health, Massachusetts, USA}
  \maketitle
} \fi

\input{2nd_version}

\end{document}

%% file: 2nd_version.tex
\if0\blind
{
  \bigskip
  \bigskip
  \bigskip
  \begin{center}
    {\LARGE\bf Confounder-Dependent Bayesian Mixture Model: Characterizing Heterogeneity of  Causal Effects in Air Pollution Epidemiology \blfootnote{The authors thank Scott Delaney, Kevin P. Josey, Fabrizia Mealli, Daniel Mork, Sally Paganin, and Massimiliano Russo for helpful suggestions and comments.  This work was partially funded by the following grants: NIH: R01ES026217, R01MD012769, R01ES028033, 1R01ES030616, 1R01AG066793, 1R01MD016054-01A1; Sloan Foundation: G-2020-13946.}}\\
     \vspace{1.5em}
   {Dafne Zorzetto$^1$, Falco J. Bargagli-Stoffi$^{2,*}$, Antonio Canale$^1$, and Francesca Dominici$^2$.}\\
\vspace{1.5em}
    {$^1$ Department of Statistics, University of Padova, Italy.\\
    $^2$ Department of Biostatistics, Harvard T.H. Chan School of Public Health, Massachusetts, USA.}\\
\vspace{1em}
    {$^*$ \texttt{fbargaglistoffi@hsph.harvard.edu}}
\end{center}
  \medskip
} \fi

\bigskip
\begin{abstract}
Several epidemiological studies have provided evidence that long-term exposure to fine particulate matter (\PMns) increases mortality risk. Furthermore, some population characteristics (e.g., age, race, and socioeconomic status) might play a crucial role in understanding vulnerability to air pollution. To inform policy, it is necessary to identify groups of the population that are more or less vulnerable to air pollution.  In causal inference literature, the Group Average Treatment Effect (GATE) is a distinctive facet of the conditional average treatment effect. This widely employed metric serves to characterize the heterogeneity of a treatment effect based on some population characteristics.
In this work, we introduce a novel Confounder-Dependent Bayesian Mixture Model (CDBMM) to characterize causal effect heterogeneity. More specifically, our method leverages the flexibility of the dependent Dirichlet process to model the distribution of the potential outcomes conditionally to the covariates and the treatment levels, thus enabling us to: (i) identify heterogeneous and mutually exclusive population groups defined by similar GATEs in a data-driven way, and (ii) estimate and characterize the causal effects within each of the identified groups. Through simulations, we demonstrate the effectiveness of our method in uncovering key insights about treatment effects heterogeneity. We apply our method to claims data from Medicare enrollees in Texas. We found six mutually exclusive groups where the causal effects of \PM on mortality are heterogeneous.
\end{abstract}

\vspace{1em}

\noindent%
{\it Keywords:}  
Bayesian nonparametric, causal inference, dependent Dirichlet process, group average treatment effect, heterogeneous causal effects.
\vfill

\newpage
\spacingset{1.9} 

\section{Introduction}
\label{sec:intro}

In the fields of health and social sciences, the identification and estimation of heterogeneous treatment effects are of paramount importance. We are particularly motivated by epidemiological studies on the effects of long-term exposure to fine particulate matter  (\PMns) on human health \citep[see][for a review]{carone2020pursuit}. 
In this context, some characteristics of the population can induce different degrees of vulnerability or resilience to air pollution. Therefore, the identification of the key factors that lead to heterogeneity of causal effect is crucial in guiding the development of environmental policies aimed at reducing the impact on the most vulnerable members of society. 
Notably, the Environmental Protection Agency (EPA) identifies race, national origin, age, sex, and/or social-economic status as potential explanatory factors  \citep{epa2022}, consistently with epidemiological studies \citep[see, e.g.,][]{ jbaily2022air}.

Bayesian nonparametric (BNP) approaches are known for their flexibility to adapt to different contexts \citep{escobar1995bayesian}. However, despite this flexibility, the literature on BNP methods for causal inference, and particularly for identifying factors that contribute to the heterogeneity of causal effects, is relatively recent \citep{linero2021and}. Researchers in this field have focused on the application or extension of the Bayesian Additive Regression Tree (BART) \citep{chipman2010bart}, and dependent Dirichlet Process (DDP) mixture models \citep{mac2000dependent, quintana2020dependent} to the causal inference framework. 

The use of BART models in causal inference was first introduced by \cite{hill2011bayesian}. Recently, \cite{hahn2020bayesian} proposed a reparameterization of the outcome model with the inclusion of the propensity score to control measured confounding further. Infinite mixture models have also been widely employed in causal inference literature. Notably, the Enriched Dirichlet process mixture model \citep{wade2014improving} and its modifications have found extensive and promising causal applications \citep{roy2018bayesian, oganisian2021bayesian}. For a review of BNP applications in causal inference, we refer to \cite{linero2021and}.

In causal inference, there is a rich literature on estimating heterogeneous causal effects via the conditional average treatment effects (CATE) \cite[see][for a review]{wendling2018comparing, dominici2020controlled}. The CATE can be specified at different levels of \textit{granularity}. For instance, at the highest level of granularity, one might estimate the treatment effect at the unit level; at a lower level of granularity, one might estimate the average treatment effect for some pre-specified \textit{groups} of the population: the group average treatment effect (GATE)\citep{jacob2019group}. Both causal estimands are special cases of CATE. 

Among the contributions, it is worth highlighting the proposal of exploring the discovery of important heterogeneous groups through an ex-post analysis method employing a fit-the-fit strategy. These techniques---proposed, for instance, by \cite{hahn2020bayesian}, \cite{krantsevich2023stochastic}, \cite{bargagli2022heterogeneous} and \cite{lee2020causal}--- utilize tree-based algorithms to identify groups characterized by heterogeneous treatment effects. Such techniques have been employed for heterogeneous group discovery in a number of relevant applications \cite[see, e.g.,][]{yeager2019national, delaney2023, bargagli2023}.

Most of the approaches for estimating the CATE, and therefore the GATE, require defining covariates values \textit{a priori}. These approaches have limitations: (i) they can be subject to the \textit{cherry-picking} problem of reporting results only for groups with extremely high/low treatment effects; (ii) they must define the groups a priori, which in turn requires a good understanding of the treatment effects, possibly from previous literature and may fail to identify unexpected, yet important, heterogeneous groups. Despite the success in accurately estimating the CATE using machine learning methods \citep{hill2011bayesian, hahn2020bayesian}, the bulk of these methods offers little guidance about which groups lead to treatment effect heterogeneity. While these papers have primarily focused on the estimation of population or sample average treatment effect, few contributions have focused on the discovery and estimation of heterogeneous groups \citep{oganisian2021bayesian, lee2020causal}.

The goal of this paper is to introduce a flexible BNP model for causal inference that can \textit{simultaneously} (i) impute the missing potential outcomes---which in turn will allow us to estimate various causal estimands of interest defined as any functions of potential outcomes and covariates---and (ii) identify mutually exclusive groups characterizing the heterogeneity in the effects. These mutually exclusive groups are identified by different group-specific causal effects (defined as GATE) and distinguished by different values of covariates.

An intrinsic feature of our proposed BNP model is that the estimation of the GATE within each population group does not depend on the \textit{a priori} choice of covariates but on the groups identified by the data. Specifically, we define the distribution for the potential outcomes, conditional on the covariates and the treatment levels, as a dependent Probit Stick-Breaking Process \citep{rodriguez2011nonparametric} mixture model. In this way, we leverage information from observable characteristics to impute missing potential outcomes \citep{holland1986statistics}. Via Monte Carlo simulations, we show that, under different scenarios of the amount of heterogeneity in the causal effects, the proposed model can (i) obtain competitive results with \citet{hill2011bayesian} and \citet{hahn2020bayesian}'s models---i.e., BART model for causal inference and Bayesian causal forest (BCF), respectively---for the estimation of average treatment effect, and simultaneously (ii) identify mutual exclusive groups and their respective GATE, accurately. In our application, we estimate the causal effect of long-term fine particle \PM exposure on mortality rate for Medicare enrollees \citep{wu2020evaluating} in Texas. We discover six mutually exclusive groups of Texas ZIP codes, characterized by different estimates of GATE and different socioeconomic and demographic features for the population.

This paper is organized as follows. In Section \ref{sec:meth}, we introduce the framework and proposed BNP model. We also derive a Gibbs sampling algorithm for sampling from the posterior distributions and estimate the model. In Section \ref{sec:simulation}, we show, via Montecarlo simulations, the ability of the proposed model to impute the missing potential outcomes and to discover heterogeneous groups. Here we also compare our model's performance with respect to the BART model, BCF, and BCF combined with CART. In Section \ref{sec:application}, we estimate the causal effects of the long-term fine particle \PM exposure and the mortality rate for Medicare enrollees \citep{wu2020evaluating} in Texas at the ZIP code level. In Section \ref{sec:discussion}, we conclude the paper with a discussion and avenues for future work.  The code for the simulations and the application can be found at \texttt{https://github.com/dafzorzetto/HTEBayes}.

\section{Confounder-Dependent Mixture Model}\label{sec:meth}

\subsection{Setup, Definitions, and Assumptions}

Let  $i$ be the study unit, with $i\in \{1,\dots,n\}$, and $T_i \in \{0,1\}$ be the binary treatment with observed value $t_i$. According to the Rubin Causal Model \citep{rubin1974estimating}, the potential outcomes for unit $i$ are defined as $\{Y_i(0), Y_i(1)\} \in \R^2$, for $i = 1, \dots, n$. Specifically, $Y_i(0)$ is the outcome when the unit $i$ is assigned to the control group, while $Y_i(1)$ is the outcome when it is assigned to the treatment group. 

In practice, however, for $i=1,\dots,n$, we observe only $y_i \in \R$, that is the realization of the random variable $Y_i$ defined as
\begin{equation*}
    Y_i : = (1-T_i) \cdot Y_i(0) + T_i \cdot Y_i(1).
\end{equation*}
Conversely, we can not observe the realization $y_i^{mis} \in \R$ of the random variable $Y_i^{mis}$ defined as $Y_i^{mis}:= T_i \cdot Y_i(0) + (1-T_i) \cdot Y_i(1)$.

Additionally, we define $\x_i \subseteq \mathcal{X}$ the $p$-dimensional vector of subject-specific background characteristics, covariates, and potential confounders---also called pre-treatment variables. Each vector $\x_i$ can contain both categorical and continuous variables. The tuple $(y_i,t_i,\x_i)$ for $i = 1, \dots, n$ therefore represents the observed quantities. 

Our goal is to estimate the causal effect of the treatment on the outcome on the basis of the potential outcome and the observed covariates $x$. To do so, it is necessary to specify causal effects of interest and the assumptions needed to identify these estimands from real-world data. In particular, the causal effects are defined as functions of the two potential outcomes. While it is possible to define the causal effect in many ways, we are particularly interested in the GATE. Generally, the CATE can be defined as
\begin{equation}\label{eq:cate}
    \tau(x):=\E[Y_i(1)-Y_i(0) \mid X_i = x].
\end{equation}
In this paper, we assume that the heterogeneity of the treatment effect is induced by a latent group structure of the observation.  Conditional to the subject-specific group label indicator  $G_i$, which in practice is never observed and will be estimated, the GATE causal estimand for group $g$ is defined as
\begin{equation}\label{eq:gate}
    \tau_g:=\E[Y_i(1)-Y_i(0) \mid G_i=g].
\end{equation}
As a specific case of CATE, GATE depends on subsets of the covariates space $\mathcal{X}$.
The group treatment effect is used to investigate and quantify the heterogeneity in the causal effects. Its estimation is useful to analyze how the effects of the treatment vary among different groups of the population.

A key goal of our proposed method is to identify the covariates $X_i$ that play a key role in the characterization of the population groups that lead to heterogeneous treatment effects.

As customary in causal inference, to identify a causal effect from the observed data, we have to make the following assumptions \citep{rubin1980randomization}. 

\vspace{0.25cm}
\noindent {\em Assumption 1: Stable Unit Treatment Value Assumption (SUTVA).}
	\begin{equation*}
		\begin{aligned}
			(i) \:\:\: &Y_i(T_i) = Y_i, \quad \mbox{for $i =1,\dots, n$};\\
			(ii) \:\:\: &Y_i(T_1, T_2, \cdots, T_i, \cdots,  T_{n}) = Y_i(T_i) \quad 
   \mbox{for $i =1,\dots, n$}.
		\end{aligned}    
	\end{equation*}
SUTVA enforces that each unit's outcome is a function of its treatment only. This is a combination of (i) consistency (no different versions of the treatment levels assigned to each unit) and (ii) no interference assumption (among the units) \citep{rubin1986comment}.

\vspace{0.25cm}
\noindent {\em Assumption 2: Strong Ignorability.} Given the observed covariate vector $x_i$, the treatment assignment is strongly ignorable if
\begin{equation*}
   \{Y_i(1), Y_i(0)\} \perp T_i \mid X_i = x_i, 
\end{equation*}
and $0<\P(T_i=1 \mid X_i=x_i)<1$, for  all $i=1, \dots, n$.
This assumption states that: (i) we have a random treatment assignment in each group conditional on some covariates values; (ii) all units have a positive chance of receiving the treatment.  

If the SUTVA and strong the ignorability assumption hold, the statistical estimand of CATE \eqref{eq:cate} can be expressed as
\begin{eqnarray}\label{eq:decomposition}
    \tau(x) &=& \E[Y_i(1) \mid X_i=x] - \E[Y_i(0) \mid X_i=x] \notag \\
    &=& \E[Y_i \mid X_i=x, T_i=1] - \E[Y_i \mid X_i=x, T_i=0];
\end{eqnarray}
and consequently also the statistical estimand of GATE \eqref{eq:gate} can be defined as:
\begin{eqnarray}
    \tau_g &=& \E[Y_i(1) \mid G_i = g] - \E[Y_i(0) \mid G_i = g] \notag \\
    &=& \E[Y_i \mid G_i = g, T_i=1] - \E[Y_i \mid G_i = g, T_i=0]. 
\end{eqnarray}
Notably, this framework can be easily extended to contexts where the treatment variable is a discrete variable with $m>2$ possible different treatments.

\subsection{Bayesian Nonparametric Model Specification}

The estimation of the causal effects can be seen as a missing data problem where, for each subject, we observe just one of the potential outcomes while the other potential outcome is always missing. Likewise, \cite{rubin1974estimating} refers to the missing potential outcomes as counterfactual outcomes. From \eqref{eq:decomposition}, we know that under strong ignorability---that is, under a sufficiently rich collection of control variables---treatment effect estimation reduces to the estimation of the conditional expectations of $\E[Y_i \mid X_i=x, T_i=1]$ and $ \E[Y_i \mid X_i=x, T_i=0]$. Provided the excellent predictive performance of Bayesian methodologies, BNP models have been widely used for this task \citep{sivaganesan2017subgroup, hill2011bayesian,roy2018bayesian,hahn2020bayesian, oganisian2021bayesian}.

Here we propose a Bayesian nonparametric (BNP) approach for the expectation of the conditional outcomes. In particular, we exploit a dependent nonparametric mixture prior---inspired by the Dependent Dirichlet process (DDP) \citep{mac2000dependent, quintana2020dependent}. Formally, we assume for each $i=1,\dots,n$:
\begin{align}
      \{Y_i \mid \x_i,t \} &\sim f^{(t)}( \cdot \mid \x_i), \notag \\ 
       f^{(t)}(\cdot \mid \x_i) &=\int_\Psi {\mathcal K}(\cdot; \psi) dG_{\x_i}^{(t)}(\psi), 
      \label{eq:model1} \\
    G_{\x_i}^{(t)} &\sim \Pi_{ x_i}^{(t)}, \notag
\end{align}
where ${\mathcal K}(\cdot;\psi)$ is a continuous density function, for every $\psi \in \Psi$, and $G_{\x_i}^{(t)}$ is a random probability measure depending on the confounders $x_i$ associated to an observation assigned to treatment level $t$. A priori we assume $G_{\x_i}^{(t)}\sim \Pi_{\x_i}^{(t)}$ where $\Pi_{\x_i}^{(t)}$ is a treatment- and confounder-dependent nonparametric process. 
Following a single-atom DDP \citep{quintana2020dependent} characterization of the random measure $G_{\x_i}^{(t)}$, we can write:
\begin{equation}
G_{\x_i}^{(t)} = \sum_{l \geq 1} \omega_{l}^{(t)}(\x_i) \delta_{\mathbb{\psi}^{(t)}_{l}},
\label{eq:asdiscreteG}
\end{equation}
where $\{\omega_{l}^{(t)}(x_i)\}_{l\geq 1}$ and $\{\psi^{(t)}_{l}\}_{l\geq 1}$ represent infinite sequences of random weights and random kernel's parameters, respectively. Notably, both random sequences depend on treatment level $t$ while the weights also depend on the confounders values $\x_i$.

Furthermore, the sequence of dependent weights is defined through a stick-breaking representation \citep{sethuraman1994constructive}, 
\begin{equation}
    \omega_{l}^{(t)}(\x_i)  = V^{(t)}_{l}(\x_i)\prod_{r<l}\{1-V_{r}^{(t)}(\x_i)\},
\label{eq:model_omega}
\end{equation}
where
$\{V_{l}^{(t)}(\x)\}_{l\geq 1}$ are $[0, 1]$-valued independent stochastic processes. The sequence of random parameters $\{\psi_{l}^{(t)}\}_{l\geq 1}$ are independent and identically distributed from a base measure $G_0^{(t)}$.

The discrete nature of the random probability measure $G_{\x_i}^{(t)}$ allows us to introduce the latent categorical variables $S_{i}^{(t)}$, that identifies the cluster allocation for each unit $i \in \{1, \dots, n\}$, whose clusters are defined by heterogeneous responses to the treatment level $t$. Assuming $\P\{S_{i}^{(t)} = l \} = \omega_{l}^{(t)} (\x_i)$, we can write model in \eqref{eq:model1}, exploiting conditioning on $S_{i}^{(t)}$, as
\begin{eqnarray}
    &\{Y_i | \x_i, t, \psi^{(t)}, S_i^{(t)} = l\}  \sim 
    {\mathcal K}( \cdot \mid\psi_{l}^{(t)}),\quad \psi_{l}^{(t)} \sim G_0^{(t)}.\notag
      \label{eq:model2} 
\end{eqnarray}
where $\psi^{(t)}$ represents the infinite sequence $\{\psi_{l}^{(t)}\}_{l\geq 1}$, for $t=\{0,1\}$.

Among the plethora of dependent nonparametric processes \citep[see the recent review by][for a detailed description]{quintana2020dependent}, we focused on the Dependent Probit Stick-Breaking (DPSB) process for its success in applications, good theoretical properties, and ease of computation \citep{rodriguez2011nonparametric}. Consistently with this, we specify: 
\begin{align}
    V_{l}^{(t)}(x_i)  = \Phi(\alpha_{l}^{(t)}(x_i)), \quad 
    \alpha_{l}^{(t)}(x_i) \sim \Norm( \beta_{0l}^{(t)} + x_i^T\beta_{l}^{(t)},1),
    \label{eq:model4}
\end{align}
where $\Phi(\cdot)$ is the Probit function and $\{\alpha_{l}^{(t)}(x_i) \}_{l\geq 1}$  has Gaussian distributions with mean a linear combination of the $p$ covariates $x_i$.

As commonly done, we assume the kernel $\mathcal{K}$ to be a Gaussian, so that model \eqref{eq:model1}--\eqref{eq:asdiscreteG} specifies to
\begin{align}
     \{Y_i | \x_i,t, S_{i}^{(t)}=l, \eta^{(t)}, \sigma^{(t)} \}\sim \Norm (\eta^{(t)}_{l},\sigma^{2(t)}_{l}).
      \label{eq:model5} 
      \end{align}
where $\eta^{(t)}$ and $\sigma^{(t)}$ represent the infinite sequences $\{\eta_{l}^{(t)}\}_{l\geq 1}$ and $\{\sigma_{l}^{(t)}\}_{l\geq 1}$, respectively. 

Prior elicitation is completed by assuming for the regression parameters  in \eqref{eq:model4} the conjugate priors
\[
\beta_{ql}^{(t)} \sim \Norm(\mu_\beta, \sigma^2_\beta),
\]
for $t=0,1$, $l\geq 1$,and  $q=0, 1, \dots,p$ and for the parameters $\eta_l^{(t)}$ and $\sigma_l^{(t)}$ in \eqref{eq:model5}
\[
 \eta_l^{(t)} \stackrel{iid}{\sim} \Norm(\mu_\eta,\sigma_\eta^2), \mbox{ and } \sigma_l^{(t)} \stackrel{iid}{\sim} \mbox{InvGamma}(\gamma_1,\gamma_2).
\]
where InvGamma($\gamma_1, \gamma_2$) represents the inverse-gamma distribution with shape parameter $\gamma_1 \in \mathbb{R}^+$ and scale parameter $\gamma_2 \in \mathbb{R}^+$, and mean equal to $\frac{\gamma_2}{\gamma_1-1}$ and variance $\frac{\gamma_2^2}{(\gamma_1-1)^2(\gamma_1-2)}$.

Sampling from the posterior distribution is straightforward via Gibbs sampling. Details are reported in the Supplementary Material.

\subsection{Groups Identification and Causal Effects Estimation}
\label{subsec:cate}
One of the advantages of BNP mixtures is their ability to cluster the observations. Consistently with our goal of defining heterogeneous causal effects, herein we discuss how to estimate mutually exclusive \textit{groups} of observations, each characterized by different GATEs.

Consistently with the BNP literature, we call \textit{clusters} the sets defining the estimated partition for each treatment level $t\in \{0,1\}$. We then combine these clusters to estimate the \textit{groups} into which the observations are divided, and for each group, we calculate a different GATE. Note that we use the term \textit{cluster} differently from \textit{group}. The former refers to the sets defined for a specific $t$ as a byproduct of the infinite mixture model specification obtained through \citet{wade2018bayesian} procedure, while the latter refers to  the final groups for which we computed different GATE.

The model specification introduced in the previous section allows us to define a group of observation based on the Cartesian product of the latent categorical variables $\{S_i^{(0)}, S_i^{(1)}\}$, for each unit $i=1,\dots,n$. Under our fully Bayesian approach $S_i=\{S_i^{(0)}, S_i^{(1)}\}$ are couples of random variables for which we can characterize the associated posterior distribution, following the computational details described in the Supplementary Material. This is customary in all Bayesian infinite mixture models, which are inherently associated with random partition models \citep{quintana2006predictive}. Under these settings, the posterior of the random partition reflects uncertainty in the clustering structure given the data \citep{wade2018bayesian}.

A general problem associated with the huge dimension of the space of partitions is the appropriate summarization of the posterior through a point estimate.  \citet{wade2018bayesian} propose a solution based on decision theory---i.e. the optimal point estimate is that which minimizes the posterior expectation of a loss function using either on Binder's loss \citep{binder1978bayesian} or Variation of Information \citep{meilua2007comparing}.

In the following analyses, we use the approach proposed by \citet{wade2018bayesian} to divide the observations into separate clusters for each value of $t \in \{0, 1\}$. By using this approach, we can associate different treatments with varying numbers of clusters, which allows for a highly flexible model. For instance, in the simulations presented in Section 3, we designed Scenario 2 to have a single cluster with a constant response for the treated subjects and different clusters with varying responses depending on the confounders for the untreated group. Our proposed approach effectively accommodates such scenarios.  Indeed we can reasonably imagine---i.e., in the context of exposure to air pollution---that people's health may be affected differently in the presence of a lower level of air pollution, instead in the presence of a high level of air pollution.

Our proposed method estimates the GATE of each group without the need for a predefined selection of a partition of the confounder space. This is achieved by directly obtaining the groups from the posterior of the model. Given that the groups are mutually exclusive, it is straightforward to identify the characteristics of the units in each group, such as the average of observed confounders or modal categories for continuous and categorical confounders, respectively.

We define the posterior distribution of the GATE for each group as the mean of the posterior distribution of the difference of the two potential outcomes for units in that group. In the following analyses, we use the mean as the posterior point estimation of GATE for each group, but other measures, such as the median, can also be used. Bayesian credible intervals for the GATE can be obtained as well. However, our proposed method allows us to define and estimate any functions of the potential outcomes conditional to the group allocation. See the Supplementary Material for an example of another causal estimand of interest for the application in Section \ref{sec:application}.

\section{Simulation Study}
\label{sec:simulation}
In this section, we present a simulation study to evaluate the performance of the proposed Confounder-Dependent Bayesian Mixture Model (CDBMM). Our objective is to investigate the model's ability to (i) accurately identify the potential outcome distributions and estimate the treatment effects, (ii) correctly identify the groups of data that describe the heterogeneity in the effects, and, as a result, accurately estimate the GATEs. To achieve this, we conduct simulations under seven different data-generating models and analyze the results to understand the model's behavior in different scenarios.

Specifically, in each scenario, we assume that the treatment variable and the confounders are Bernoulli random variables, with different probabilities of success. Each scenario assumes a different conformation of the groups, where the units are allocated according to covariates values. Conditionally on group allocation we simulate the potential outcomes.

In the simulated scenarios, we investigate different situations that can arise in real data applications: when the expected value of the outcome decreases with the treatment, but the intensity of this decrease varies across different groups (Scenarios 1, 3, and 6, with different degree of difficulty for the groups identification); a setting where the population has different outcomes under the control group but similar outcomes under treatment (Scenario 2); the combination of the two previous situations (Scenarios 4 and 5, with more groups and confounders); and the case of homogeneous treatment effects where we have only one group (Scenario 7). 
Refer to Appendix B in the Supplementary Materials for the specific details of the seven simulated scenarios and respective visualization of the distributions of the causal effects. We choose the hyperparameters such that the prior is non-informative and in common for all the settings (details in the Supplementary material).

The performance of the proposed approach is compared to those obtained with the BART model by \citet{hill2011bayesian}---using the \texttt{R} package \texttt{bartCause}---and with the BCF approach by \citet{hahn2020bayesian}---using the package \texttt{bcf} available in GitHub. We chose BART and BCF since benchmarks as these models have shown particular flexibility and an excellent performance---with the need of no or little hyper-parameter tuning---in both prediction tasks \citep{linero2018bayesian1, hernandez2018bayesian} and in causal inference applications \citep{hill2011bayesian, hahn2020bayesian}. Also, results from Monte Carlo simulations studies show an excellent performance of BART and BCF in causal inference settings \citep{dorie2019automated}. 
BART and BCF do not have a direct group-based characterization of the heterogeneity of the causal effect, but the groups can be obtained with second-step, where the units are grouped according to their causal effects at unit level by classification and regression trees (CART), introduced by \citet{breiman1984cart}. In particular, the benchmark for group identification is BCF + CART combo, where we use the \texttt{R} package \texttt{rpart} for the CART.

First, we analyze the result for the ATE. As illustrated in Figure 1, the results obtained from the three models are quite similar in terms of both bias and mean square error. In particular, the bias is close to zero, as reported in the first line of boxplots of Figure 1, where the median of the boxplots is close to the red horizontal line that indicates a bias of zero; and the variability among the censorious is small and correlated to the variability of the simulated data---e.g. the scenario 7 has boxplot with longer tails that the other scenarios, due to a bigger simulated values for the parameters $\sigma^{(0)}$ and $\sigma^{(1)}$. The mean square errors, in the second line of boxplots in Figure 1, reflect the simulated variability in each of the seven scenarios. The MSE results for the three models are comparable, with smaller median values for CDBMM, even though, there are some outliers for scenario 4 for this model.  

\begin{figure}
\begin{center}
\includegraphics[width=6in]{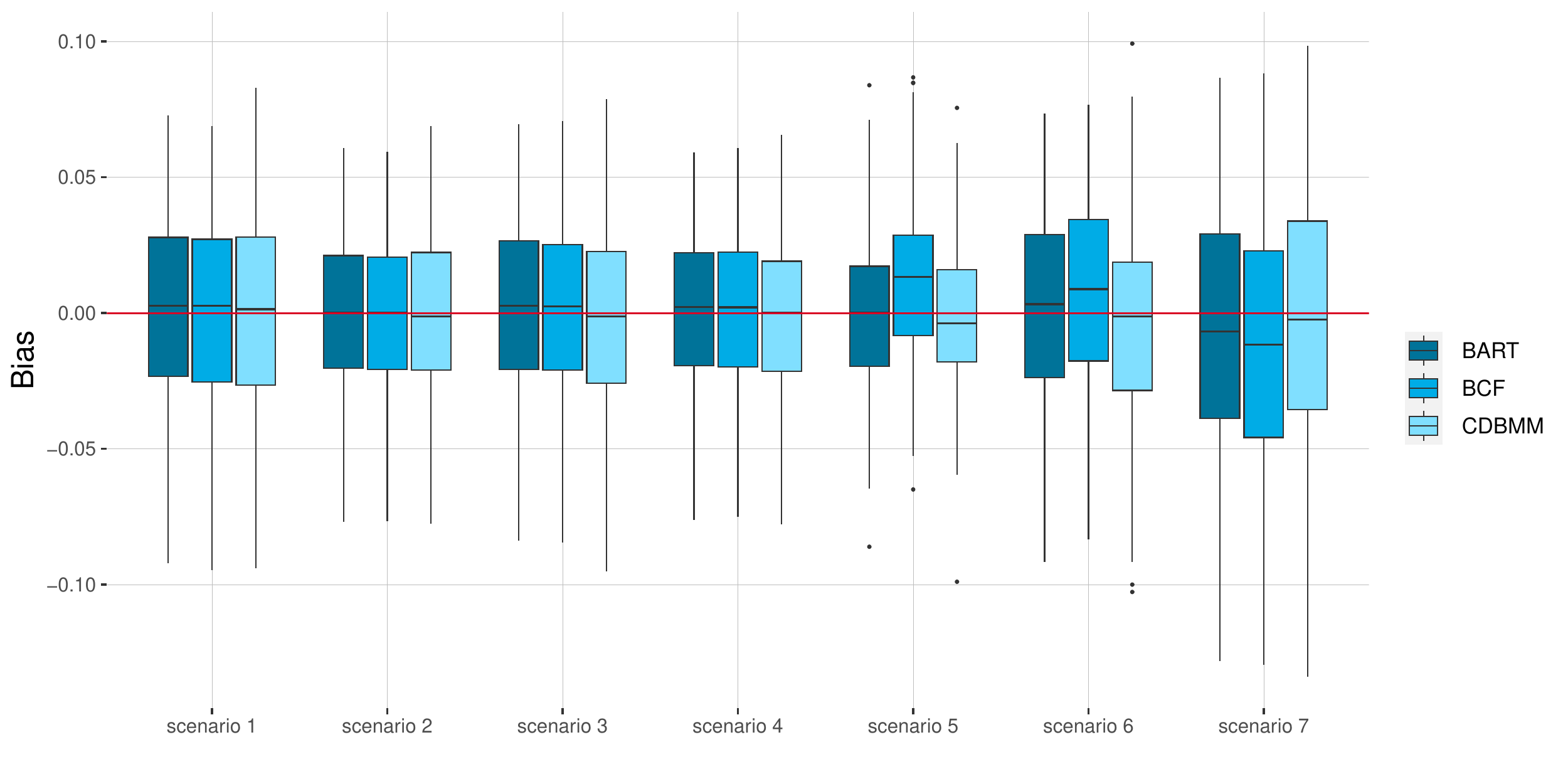}\\ \includegraphics[width=6in]{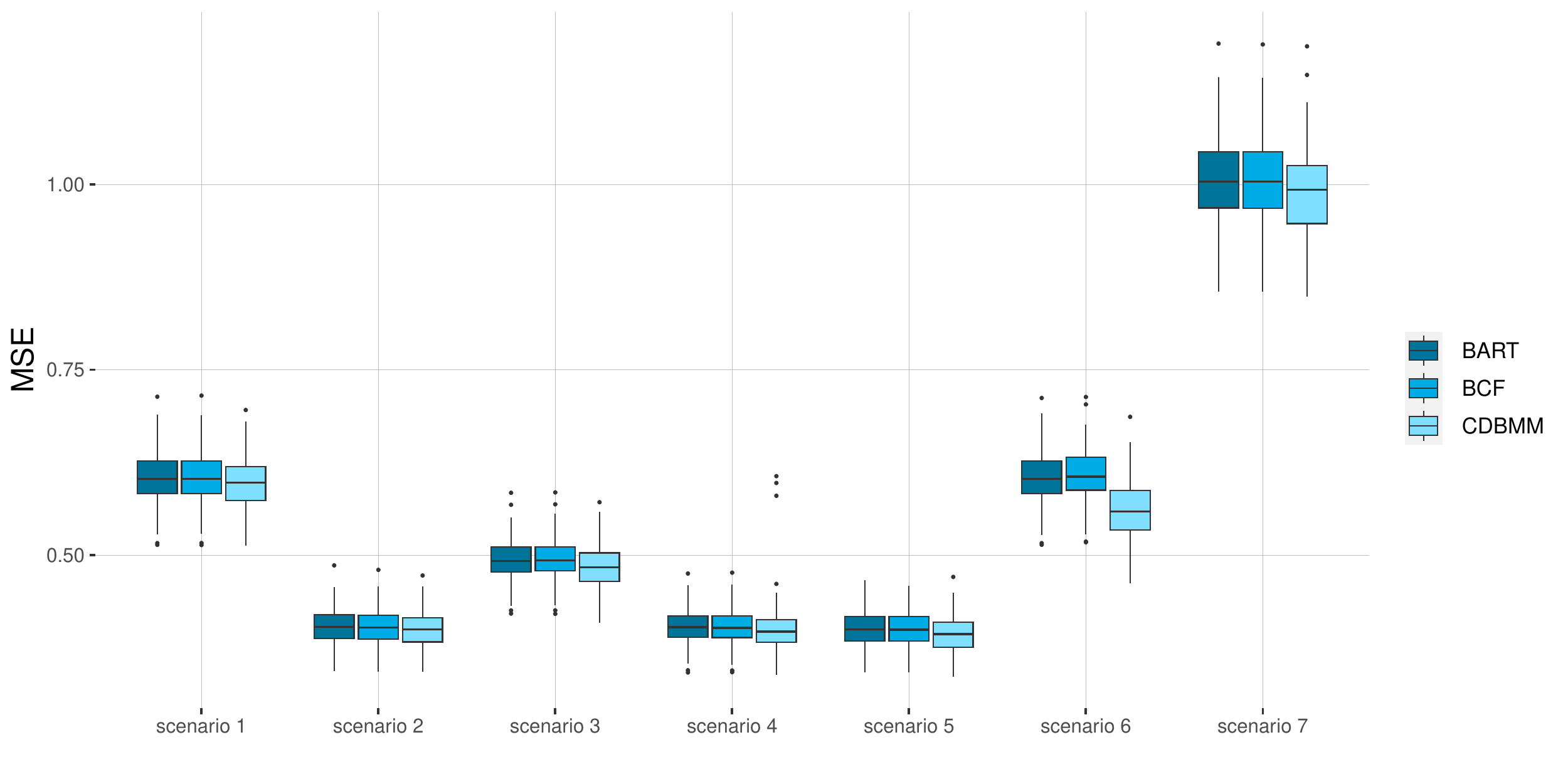}
\end{center}
\caption{Comparison of estimation of average treatment effects between CDBMM, BART, and BCF: bias and mean square error.  \label{fig:sim_bart}}
\end{figure}

The proposed method not only produces accurate ATE estimates but also excels in estimating the GATEs. In section C in the Supplementary Material we have reported the visualization of the estimated GATEs for each group in the seven simulated scenarios. 

The ability of CDBMM to estimate the GATEs is a combination of accurate estimation of the distributions of the potential outcomes and accurate group identification. 
To quantify the model's capability of identifying the exact number of groups and correctly allocate them the units we use the adjusted Rand index (ARI)---using the \texttt{R} package \texttt{mcclust}. ARI is a cluster comparison measure, that informs of the goodness of the estimated partition, compared with the simulated partition. It takes values in $[0,1]$, with $0$ indicating that the two partitions do not agree on any pair of units and $1$ indicating that the partitions perfectly match. We utilize as a benchmark the BCF + CART combo, and the compared results are reported in Table 1. The proposed method has superior performance, in mean and variability, broadly concerning the BCF+CART combo in Scenarios 1--6 where heterogeneity is present.  In the case of homogeneity (Scenario 7), the methods are comparable as both methods correctly find a single group.

\begin{table}
\centering
\begin{tabular}{ccccccc}
\hline
 && \multicolumn{2}{c}{CDBMM} && \multicolumn{2}{c}{BCF + CART} \\
    && mean  &  sd && mean & sd \\ 
\hline

  \rowcolor{lightGray}
    Scenario 1 &&  0.9995 & 0.0016 && 0.8203 & 0.0185 \\ 
\rowcolor{Gray2}
  Scenario 2 && 0.9910 & 0.0390 &&  0.8203 & 0.0185 \\ 
  \rowcolor{lightGray}
  Scenario 3 &&  0.9981 & 0.0046 && 0.8158 & 0.0416 \\
\rowcolor{Gray2}
  Scenario 4 &&  0.9926 & 0.0320 && 1.0000 & 0.0000 \\ 
  \rowcolor{lightGray}
  Scenario 5 && 0.9905 & 0.0199 && 0.7935 & 0.0283 \\
\rowcolor{Gray2}
  Scenario 6 && 0.9757 & 0.0425 && 0.7351 & 0.1341 \\ 
  \rowcolor{lightGray}
  Scenario 7 && 1.0000 & 0.0000 && 1.0000 & 0.0000 \\
\hline
\end{tabular}
\caption{Mean and empirical standard deviation of the adjusted Rand index computed for the seven settings with CDBMM and BCF+CART combo.}
\label{table:rand_p1}
\end{table}

In the Supplementary Materials, we have investigated the sensitivity analysis and the run time comparison among CDBMM, BART, and BCF for different sample sizes, respectively reported in Section D e Section E.

\section{Heterogeneous Effects of Air Pollution Exposure on Mortality}
\label{sec:application}
The literature has found significant evidence of decreasing mortality deriving from long-term exposure to lower levels of \PM \citep[see][]{carone2020pursuit, wu2020evaluating}. While previous studies have made significant contributions in assessing the average treatment effect of long-term \PMns, they have primarily done so without considering potential heterogeneity in the causal effects. However, understanding how the causal effect can vary within different groups of individuals is crucial in health studies, as it can inform the development of more effective health policies.

With a focus on uncovering vulnerability/resilience in the causal effects in the context of an environmental study, we apply the proposed model to discover the heterogeneity in the health effects of exposure to higher levels of air pollution in Texas for the elderly population. Texas is a crucial case study for understanding air pollution vulnerability because of its unique demographic makeup and exposure disparities. First, recent literature has shown that, in Texas, black and low-income groups are exposed to higher levels of air pollution, whereas college graduates and high-income groups are exposed to lower levels \citet{li2019racial}. Second, Texas has a high proportion of Hispanic residents, which makes it a valuable case study for examining the health impacts of air pollution on this demographic group. According to the US Census Bureau, Hispanic residents make up over 39\% of the Texas population \citep{UScensus}. Studies have shown that low-income and Hispanic communities are more likely to be exposed to higher levels of \PM compared to wealthier and non-Hispanic communities \cite{jbaily2022air}.

Using the information on Texan Medicare enrollees (i.e., individuals older than 65), we investigate the heterogeneous causal link between long-term \PM exposure and mortality. Our analysis depicts how our method can discover mutually exclusive groups, estimate the heterogeneity in the effects of long-term exposure to \PM on the mortality rate, and identify the social-economical characteristics that distinguish the different groups.

Understanding the variations in causal effects within different groups of individuals holds paramount importance in the realm of health studies. Such insights can serve as vital building blocks for the development of more effective and targeted health policies. Our method, CDBMM, has a distinct capability to precisely target this type of estimand (namely, the GATE) by producing a flexible derivation of the heterogeneous effect groups through a data-driven method and the estimation of the causal effects of those groups. By leveraging CDBMM, we uncover these heterogeneous groups and estimate the group-specific treatment effects, providing policymakers with invaluable information on the vulnerability/resilience of specific groups of the populations.

\subsection{Data}

We conduct our analysis at the ZIP code level (1,929 units), where we have data on the following variables: the average \PM levels during the years 2010 and 2011; the mortality rate in the 5 follow-up years; census variables such as the percentage of residents for different races/ethnicities (in particular, categorized as Hispanics, blacks, whites, and other races); the age of each Medicare enrollee ($\leq 65$ years of age) and their sex (female/male); the average household income; the average home value; the proportion of residents in poverty; the proportion of residents with a high school diploma; the population density; the proportion of residents that own their house; the average body mass index; the smoking rate; the percentage of people who are eligible for Medicare (this variable is a proxy of low social-economic status and is reported as S.E.S.). Moreover, we also have access to meteorological variables: the averages of maximum daily temperatures and the relative average of humidity during summer (June to September) and winter (December to February). 

The distribution of the population in Texas during 2010, represented in map (a) in Figure 2, is clearly concentrated around the main cities, such as Dallas, San Antonio, Austin, and Houston, while expansive areas are quite empty, due to the desert ecosystem. Consequently, we consider only the ZIP codes with a population density different from zero and more than 10 Medicare enrollees, such that we have enough records in the Medicare dataset for those ZIP codes. For these ZIP codes, the observed values of \PM and of mortality rates per 100,000 in Texas are illustrated in Figure 2---(b) and (c), respectively. Not surprisingly, the highest values on record for \PM are primarily concentrated in the urban areas, while the mortality rate doesn't show any particular pattern.

\begin{figure*}
\begin{center}
\includegraphics[trim={4cm 7cm 0cm 5cm}, width=6in]{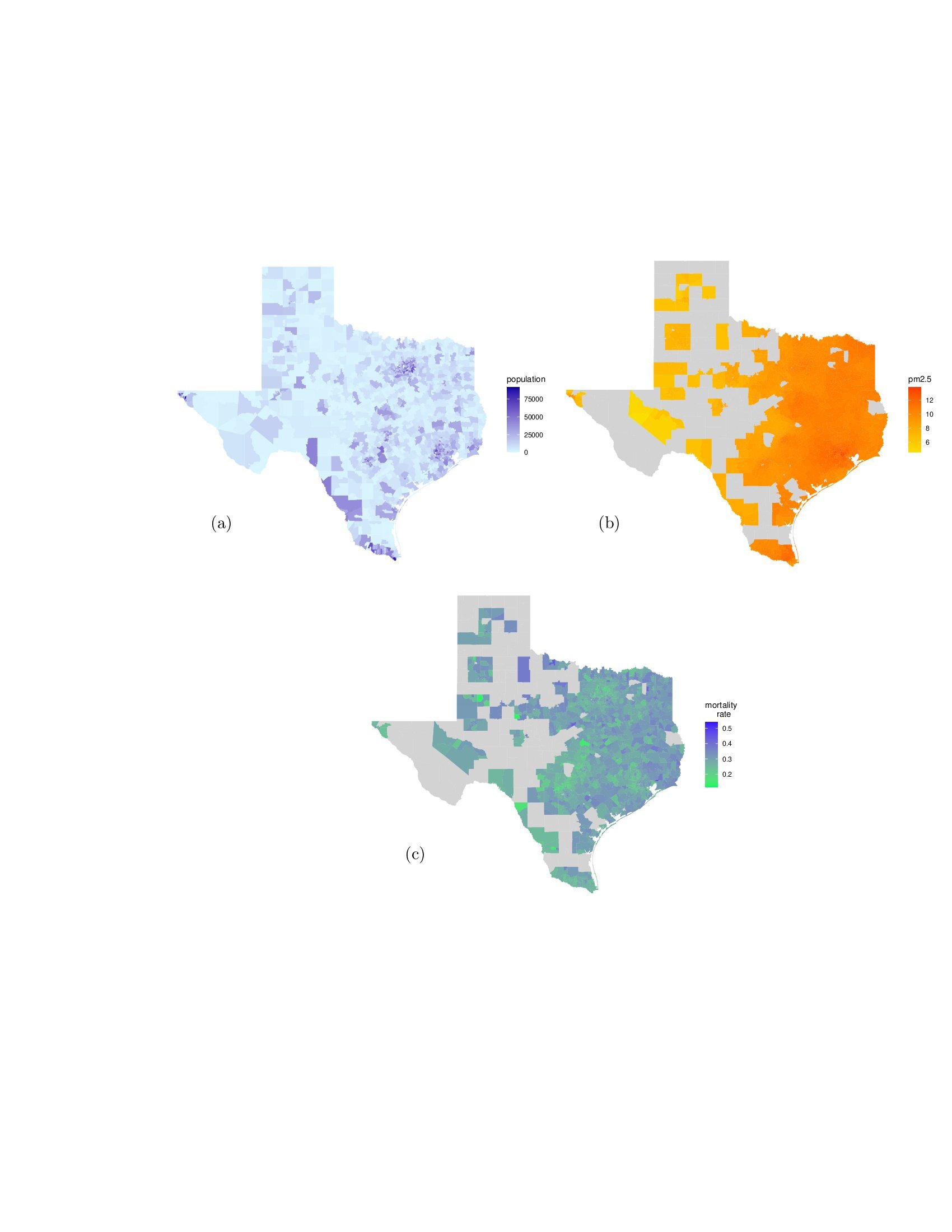}
\end{center}
\caption{(a) Population density during 2010 in Texas (USA) (b) Average of long-term \PM exposure in 2010-2011 in Texas. The data are aggregated by ZIP codes. (c) The mortality rate at 5 follow-up years for each Texan ZIP code (rate per $100,000$). The gray areas indicate the ZIP codes with population density different from zero and more than 10 Medicare enrollees.  \label{fig:map_ca}}
\end{figure*}

\subsection{Results}

We define the treatment variable as $T=1$ if the average \PM in 2010 and 2011 is above the threshold of $10 \mu g/m^3$ and $T=0$ otherwise. The choice of $10 \mu g/m^3$ as a threshold aligns with the new proposal for the National Ambient Air Quality Standard (NAAQS) established by the Environmental Protection Agency (EPA) \citep{epa2022}. On January 27, 2023, the EPA announced a proposal to lower the annual to reduce the annual PM2.5 National Ambient Air Quality Standard (NAAQS) in a range of 9 to 10 $\mu g/m^3$ \citep{epa2022}. To provide relevant information to current EPA regulatory decision-making we started exploring the potential harm associated with exposure to PM$_{2.5}$ levels surpassing the new proposed threshold for NAAQS. Thus, in our application, we dichotomize the exposure to PM$_{2.5}$ above and below 10 $\mu g/m^3$. In this specific context, our central objective differs from the estimation of the causal effect of continuous exposure on the outcome, which has been extensively investigated in prior studies \cite[see, e.g.,][]{josey2023air}. Our approach underscores that, although the exposure variable inherently maintains its continuous nature, we treat it as binary to effectively address this critical policy question.
See the Supplementary Material for additional details about the threshold and study design.

CDBMM identifies six mutually exclusive groups in the matched ZIP codes: four where exposure to higher levels of \PM increases the mortality rate, and two where exposure to higher levels of \PM decreases the mortality rate. Figure 3 presents the posterior distribution of the GATEs for each identified group. The vertical black line represents the null causal effect, which is indicated by a GATEs equal to 0. In the Supplementary material, we have also defined and estimated the group average risk ratio (GARR). GARR provides complementary information to GATE, and their combination furnishes a deeper insight into the heterogeneity in the causal effects in the case of our application. 

The four groups identified by CDBMM where exposure to higher levels of \PM increases the mortality rate have positive GATE values (in order from highest to lowest: (f) 0.143, (e) 0.097, (d) 0.039, (c) 0.006). While the groups where exposure to higher levels of \PM decreases the mortality rate have negative GATE values (in order from lowest to highest: (a) -0.040 and (b) -0.007). 

The majority of the population ($90\%$ of ZIP codes) is included in the groups (b), (c), and (d). In the bigger group (c)---including $45,4\%$ of the ZIP codes--- the mortality rate of the population increases by $0.006$ with respect to the mortality rate under a lower level of \PMns; the group (d)---including $13\%$ of the ZIP codes---depicts an increment of $0.039$ of the mortality rate. Conversely, group (b), including $30,5\%$ of the ZIP codes, has a decrement of mortality by $0.007$. Three small groups (a), (e), and (f) are also discovered ($10\%$ of ZIP codes). More extreme effects characterize these groups. An increment of up to $0.143$ of the mortality rate when the sub-population in (f) is exposed to a high level of \PM and an increment of $0.097$ for the group (e), while the group (a) has a decrement of $0.04$. The percentages of ZIP codes allocated in the various groups are reported in Figure 4.

\begin{figure}
\begin{center}
\includegraphics[width=4.5in]{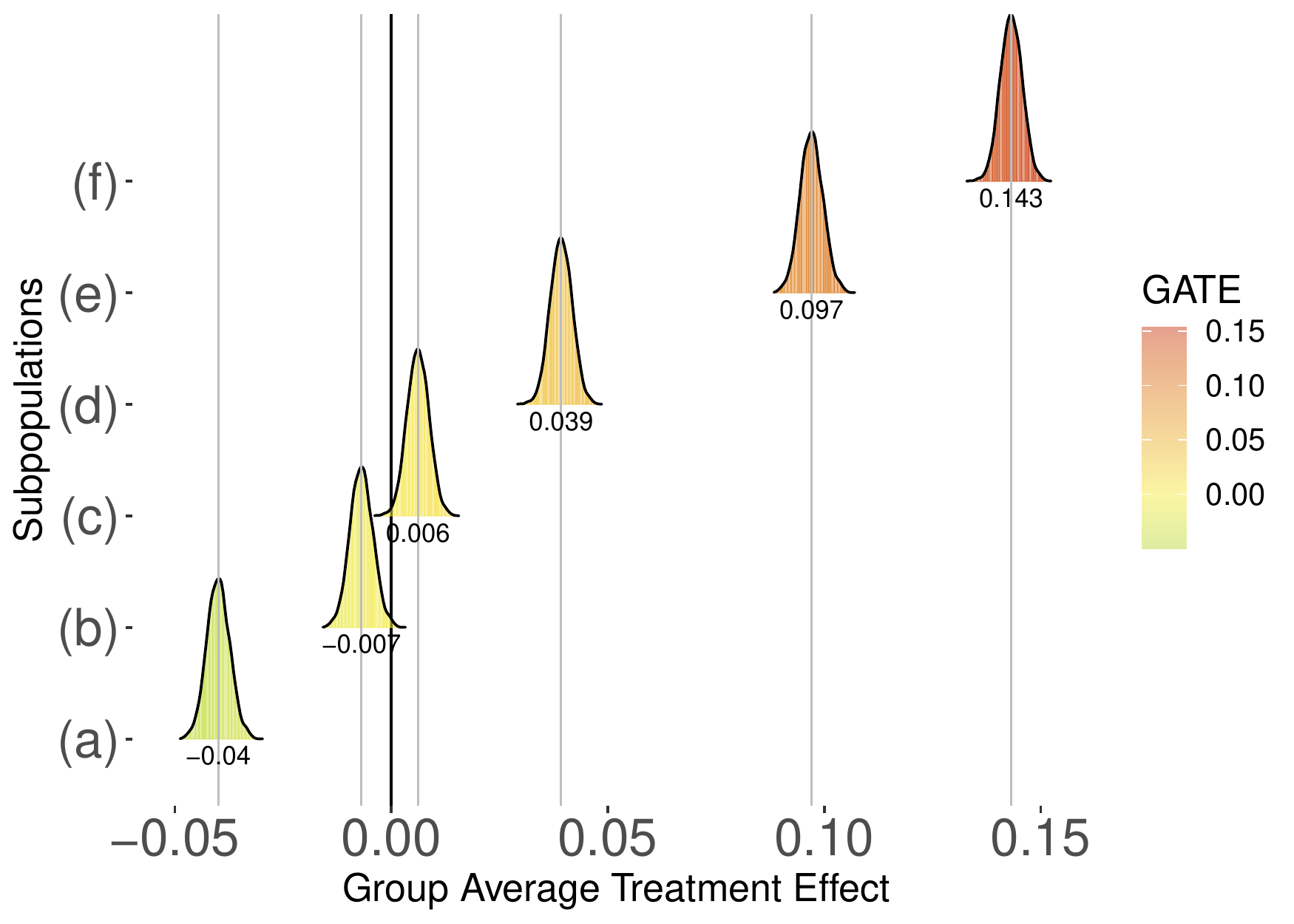}
\end{center}
\caption{Posterior distribution of GATEs for the six estimated groups in the ZIP codes. The black line identifies the null causal effects and the gray lines are the mean of each posterior distribution for the GATEs.  \label{fig:post_CE}}
\end{figure}

The demographic and socio-economic characteristics of the ZIP codes are helpful in understanding the differences in the causal effects in each identified group. The estimated model identifies mutually exclusive groups. Therefore, we can describe the distribution of these characteristics for the ZIP codes allocated in the different groups. 

In the spider plots reported in Figure 4, we can observe, for each group, the different distribution of the variables sex (close to the center indicates a higher percentage of women in the population of the ZIP codes, far to the center a higher percentage of men), percentage of white, Hispanic, black and other race (where smaller percentages are closer to the center), old (where the ZIP codes with age mean close to 65 years for Medicare enrollees are close to the center, and older population far from the center), and poor (the center of the spider-plot indicates a population with high income---i.e., a lower percentage of dually eligible individuals as proxy---, and far from the center lower income). Each group is identified with a different color, while the grey area reports the mean of these variables among all the analyzed ZIP codes after matching.

The groups where exposure to higher levels of \PM increases the mortality rate are characterized by poorer populations for (c) and (d) and a higher percentage of black and/or other races for all four groups. Consistently with the intuition, these groups are characterized by higher percentages of people from minority groups for long-term exposure to \PM. This could be likely due to the fact that minorities, often associated with low income, are structurally exposed to higher levels of air pollution. Exposure effects might accumulate over time---as is likely the case with \PM---leading to increased mortality rates \citep[see, e.g.,][]{jbaily2022air}. The groups (e) and (f) are characterized by a young (close to 65  years) and rich population compared to the mean among all the analyzed ZIP codes, specifically with white and Hispanic women in group (e) and male for group (f).

In juxtaposition, the groups (a) and (b) where exposure to higher levels of \PM decreases the mortality rate is mainly composed of a population with a higher percentage of Hispanics compared to the mean among all the analyzed ZIP codes and the other identified groups. In particular, the group (a) is also composed of a big community of blacks and other races.
This decrease in mortality when being exposed to higher levels of air pollution, while surprising, has already been documented in the literature \citep{jbaily2022air}. This finds an explanation in potential survival bias \citep[see, e.g.,][]{shaw2021evaluation}. Survival bias happens in cohort studies that start later, leading to the most vulnerable individuals in certain groups dying before entering the cohort. In this case, the individuals entering the cohort are the most resilient ones and might depict a decreasing mortality effect even when exposed to higher levels of pollutants. This is likely to be the case for these two groups.

In the Supplementary Materials, we investigate the spatial distribution of the discovered cluster in Texas. 

\begin{figure*}
\begin{center}
\includegraphics[trim={2cm 2.5cm 2cm 2.5cm}, width=1.8in]{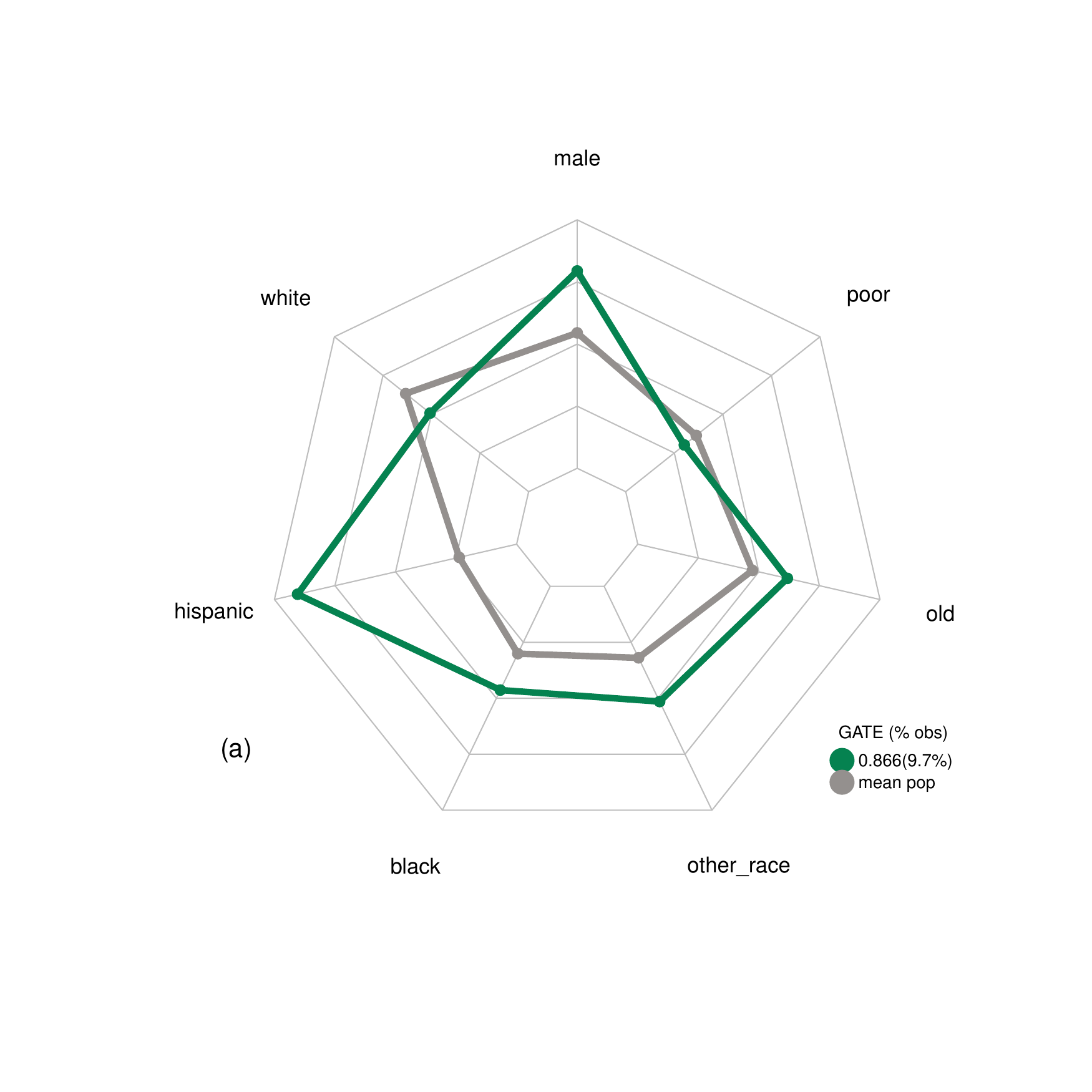}\;\includegraphics[trim={2cm 2.5cm 2cm 2.5cm},width=1.8in]{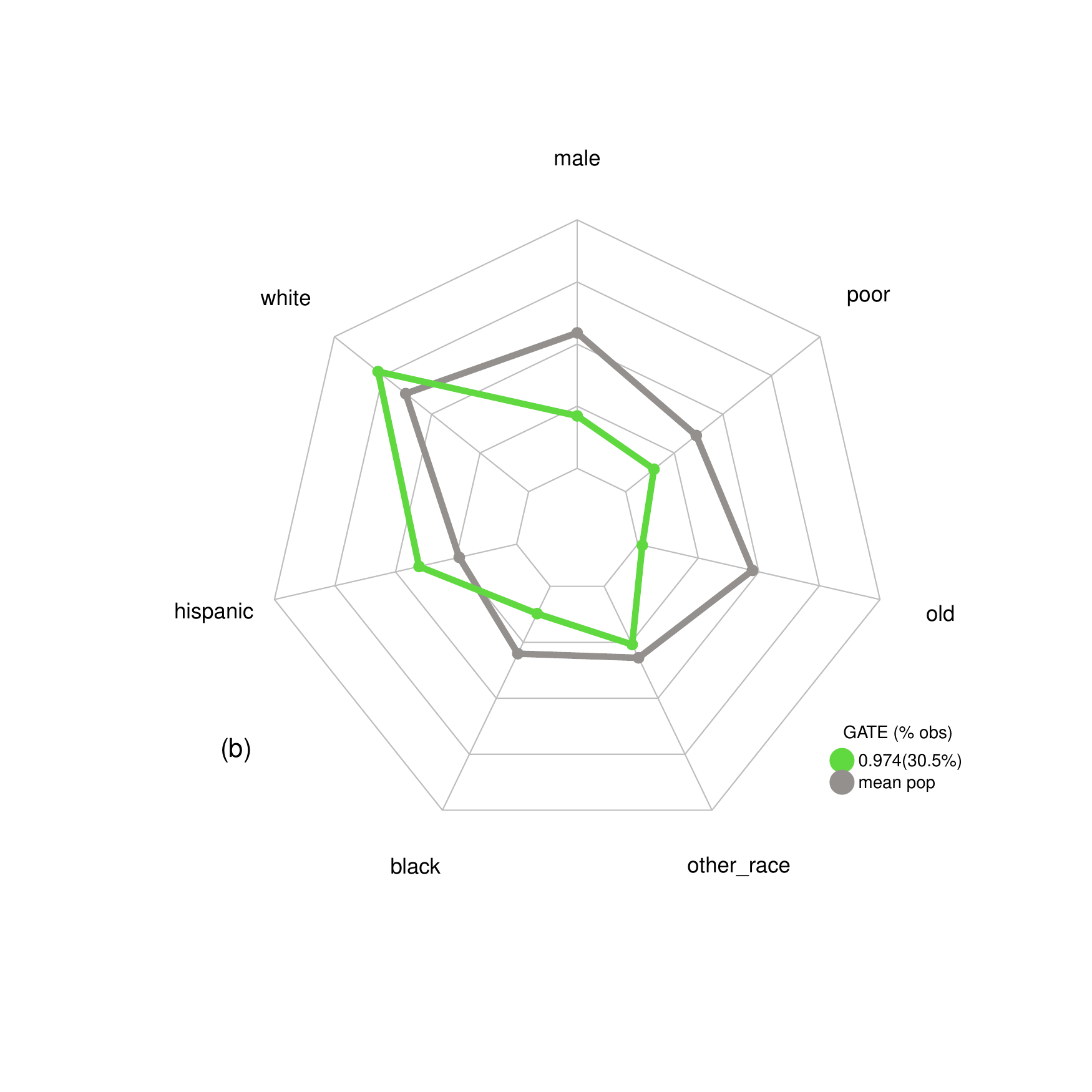}\; \includegraphics[trim={2cm 2.5cm 2cm 2.5cm},width=1.8in]{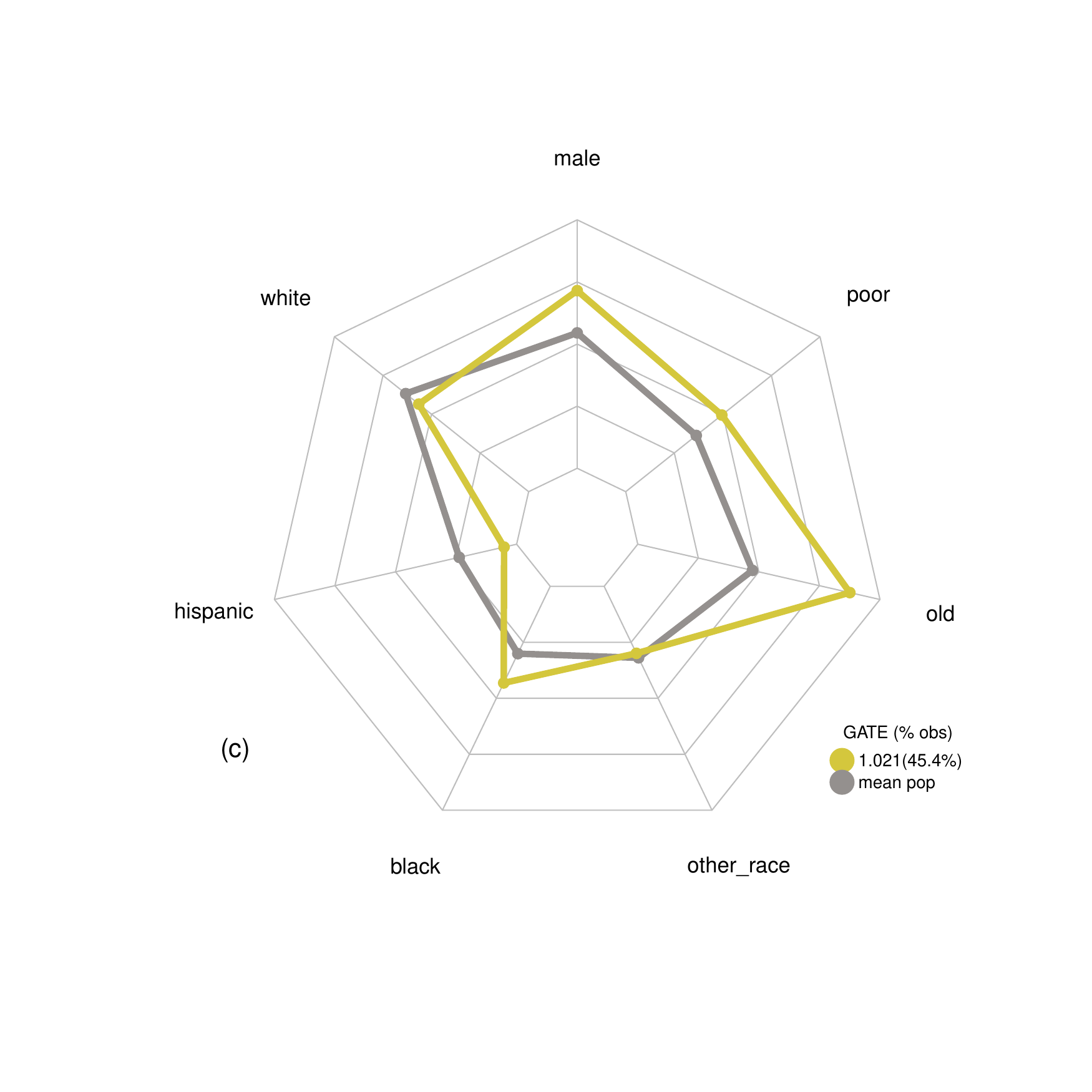}\\ \includegraphics[trim={2cm 2.5cm 2cm 2.5cm},width=1.8in]{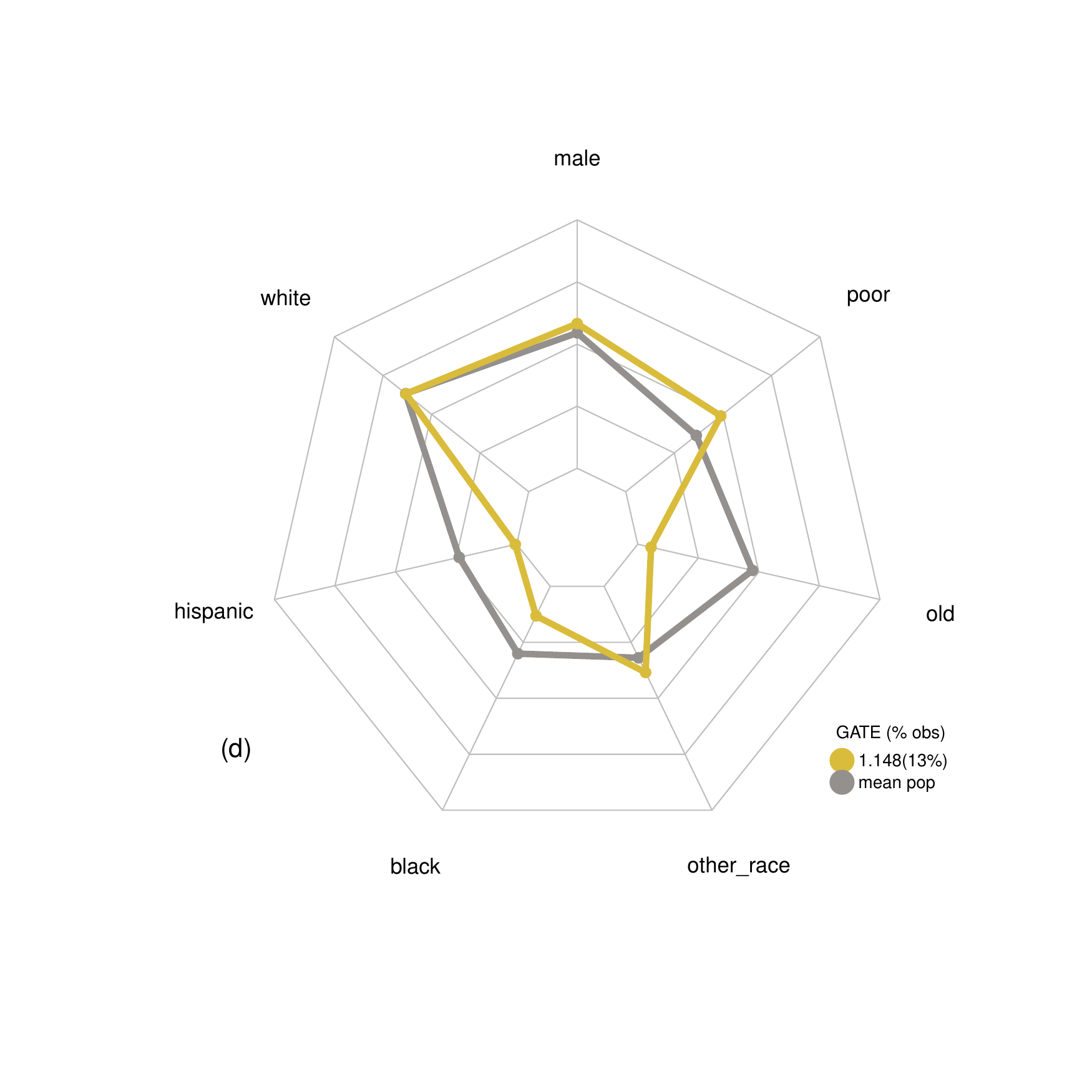}\;\includegraphics[trim={2cm 2.5cm 2cm 2.5cm}, width=1.8in]{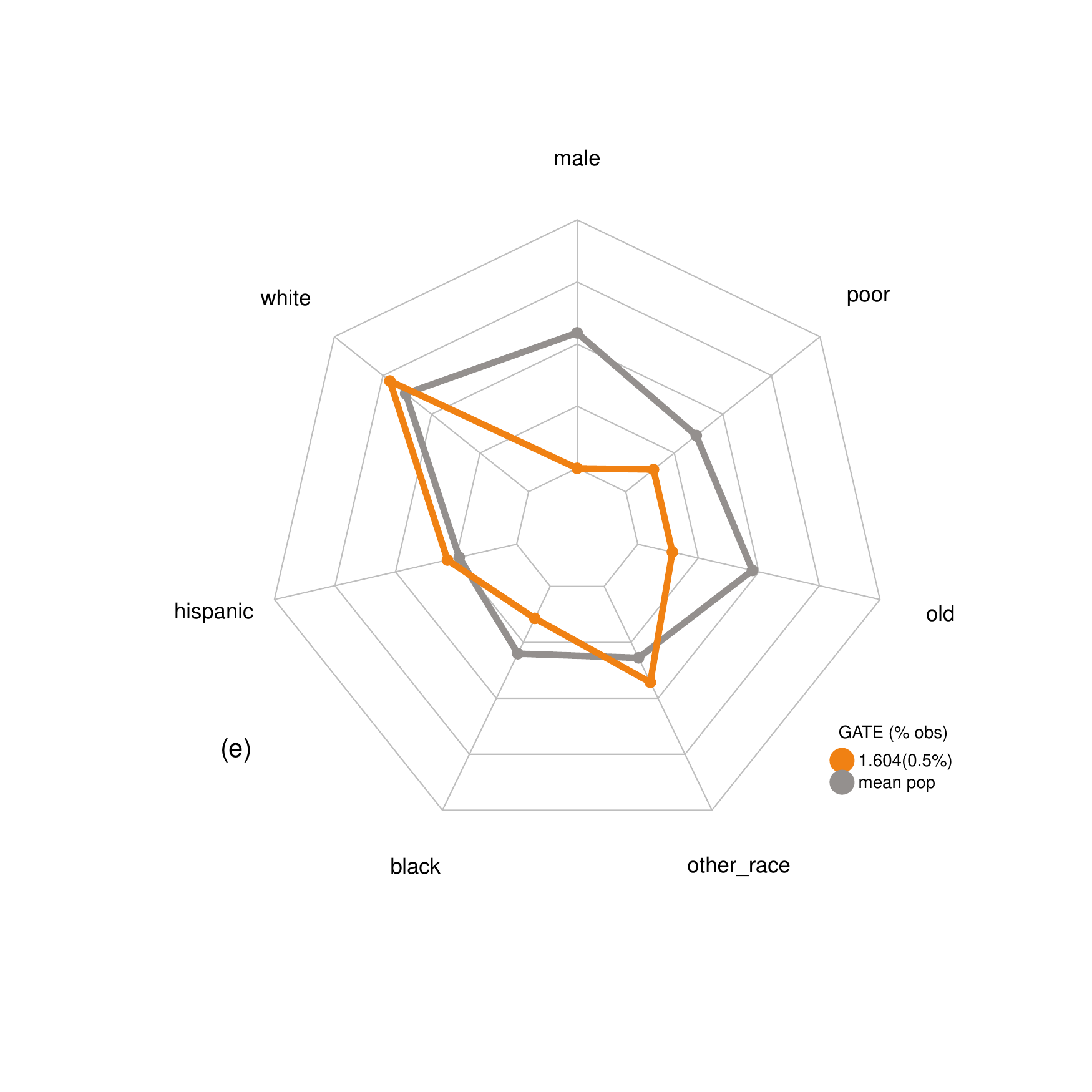}\; \includegraphics[trim={2cm 2.5cm 2cm 2.5cm}, width=1.8in]{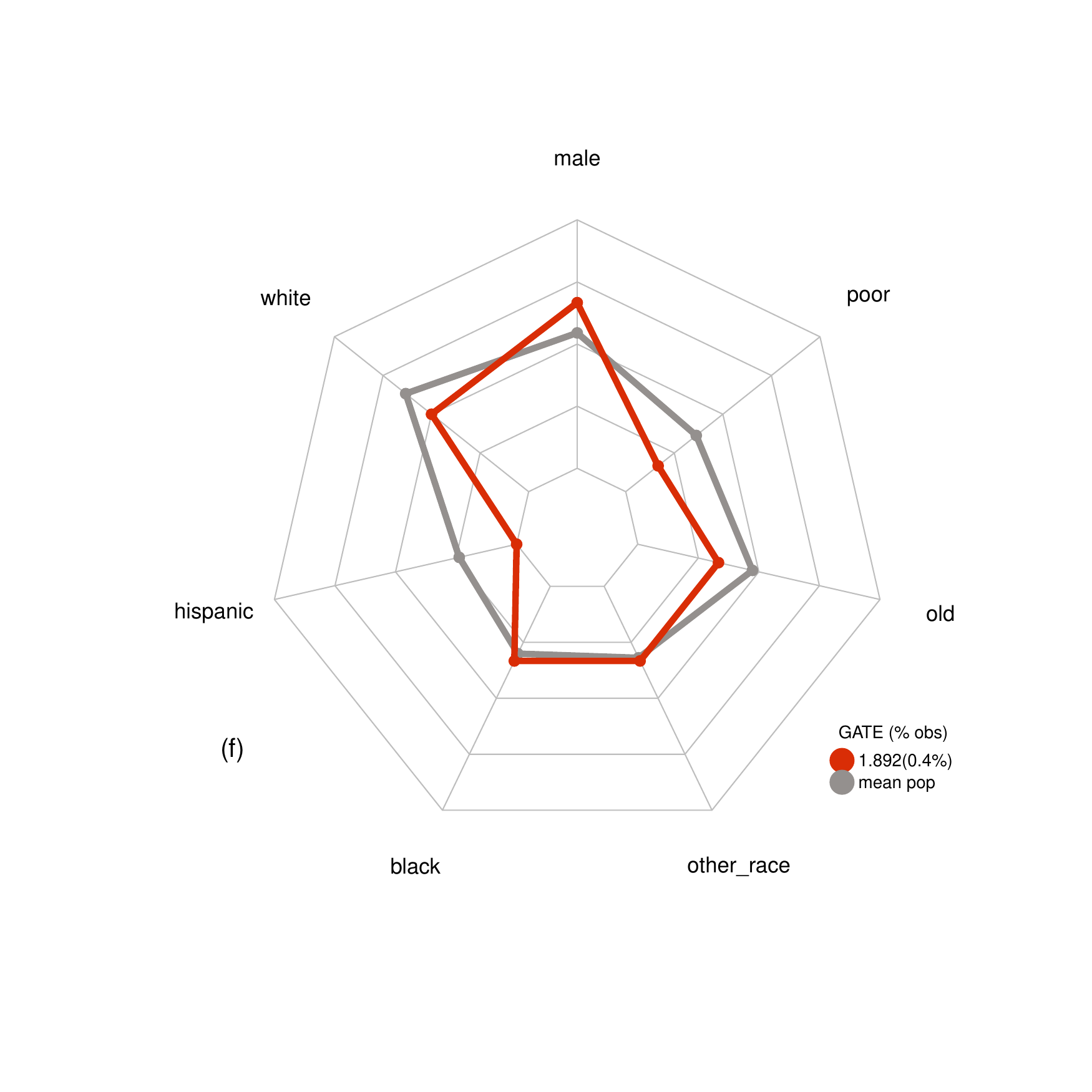}
\end{center}
\caption{Representation of the characteristics of the identified groups. Each spider plot reports in the colored area the group-specific characteristics---the mean of the analysed covariates---and in the gray area the collective characteristics---the mean of the covariates among all the analyzed Texan ZIP codes. We can consider the gray area as the benchmark to understand how the characteristics of each group differ from the collective characteristics of the analyzed Medicare enrollees in Texas. \label{fig:covariates}}
\end{figure*}

\section{Discussion}
\label{sec:discussion}
In this paper, we introduce a novel approach (CDBMM) for estimating GATEs under a Bayesian nonparametric framework. We find that the proposed CDBMM is a flexible approach, capable of identifying groups that drive heterogeneity of the causal effects. A key feature of the CDBMM is that the model can reliably identify the groups that lead to heterogeneity and estimate the causal treatment effects within each group. In other words, under our proposed approach, we do not need to make choices a priori regarding which covariates could be the key drivers of heterogeneity. 

The proposed CDBMM method offers a unique capability to \textit{simultaneously} identify distinct groups with similar treatment effects within the group and also estimate group specific causal effects. In contrast, existing methods like BART and BCF excel at estimating heterogeneous treatment effects but were not originally designed to handle both 1) discovery of heterogenous groups; and 2) group specific estimation of causal effects  at the same time. CDBMM seamlessly integrates group discovery and treatment effect estimation within its mixture specification, eliminating the need for a separate post-processing step. While post-processing techniques like BCF+CART \citep{hahn2020bayesian} demonstrate good performance in heterogeneous group discovery \cite[see, e.g.,][]{bargagli2022heterogeneous}, the results of our simulations suggest they slightly underperform CDBMM in correctly identifying true groups (Section 3). This is possibly due to the fact that CDBMM explicitly targets both tasks without requiring extra post-processing. Additionally, post-processing-based methods often require fine-tuning of multiple hyperparameters, unlike CDBMM, which operates without such adjustments. Finally, CDBMM introduces a propagation of uncertainty that method combinations tend to overlook (for instance, the majority of tree-based post-processing techniques typically do not account for the uncertainty associated with subgroup discovery through CART), while also facilitating consistent uncertainty quantification in the estimation of GATE under a proper Bayesian setting.

In our simulation studies, we find that CDBMM is competitive with BART and BCF in terms of estimating ATE. Additionally, CDBMM is able to correctly identify the groups, competitively with BCF + CART combo, and estimate the GATE with a high degree of accuracy. Our application to long-term exposure to \PM effects in Texas reveals six distinct groups that characterize the heterogeneity in treatment effects. The groups identify the key factors in the population characteristics that explain different degrees of vulnerability or resilience to air pollution. In particular, we find that the groups characterized by male black/other races with low incomes or a younger and richer population were found to have an increased mortality rate from exposure to \PM. This finding is consistent with previous literature \citep[see, e.g.,][]{jbaily2022air}. Conversely, for groups characterized by a high percentage of Hispanics, we observe a decrease in mortality which might be due to potential survival bias.

Since our applied objective is detecting vulnerability and resilience to air pollution exposure and accurately estimating treatment effects within these vulnerable and resilient groups, the capability of CDBMM to simultaneously identify heterogeneous treatment effects and estimate these effects without the need for postprocessing method proves essential for our investigation and potentially guiding the development of environmental policies aimed at the most vulnerable groups.
 
Furthermore, using the CDBMM's approach ensures that we obtain sound inference and uncertainty characterization for each of the identified groups which wouldn't be guaranteed with postprocessing methods that usually disregard uncertainty propagation. Finally, despite our dataset's substantial size, consisting of 1,400 observations, we find that CDBMM's computational requirements are not significantly greater than those of alternative methods, further highlighting its practicality in our research context. Moving forward, further research is needed to address our applied research question at a U.S. national level.

While our method is been defined for a binary treatment, it can be easily extended to handle categorical treatment as well. Similar approaches can be used in principal stratification and mediation analysis settings where one can assume the presence of a mediator/intermediate/post-treatment variable between the treatment and the outcome of interest. In such scenarios, CDBMM flexibility and performance in heterogeneous group discovery could provide an extremely valuable tool.

In conclusion, the proposed method shows promising results in the flexibility and efficiency of the nonparametric potential outcomes distribution and GATE estimation. In particular, the ability to identify groups defined by similar group-specific causal effects and the non-dependence of prior assumptions make the method versatile for practical applications. Furthermore, the extension to other causal inference settings paves the way for future research.

\bibliographystyle{chicago}
\bibliography{bib_2}

\pagebreak
\appendix

\counterwithin{figure}{section}
\counterwithin{table}{section}
 \pagenumbering{arabic}
    \setcounter{page}{1}
    
\begin{center}
    \LARGE \textbf{SUPPLEMENTARY MATERIAL}
\end{center}

\section{Posterior computation}
\label{sec:app_gibbs}

\citet{rodriguez2011nonparametric} proves that the finite truncation of the DPSB process is a good approximation; therefore, we can rewrite Equation 5 as a finite mixture to $L<\infty$ components with $L$ a reasonable conservative upper bound. \citet{rodriguez2011nonparametric}'s proof is a key point that allows us to provide a simpler algorithm without losing the robustness of the model. 

The Gibbs sampling algorithm for model fitting, which we define below, is inspired by the algorithm proposed by \citet{rodriguez2011nonparametric} to obtain draws from the posterior distribution. Following the steps in the algorithm~\ref{alg:gibbs}, in each iteration $r=1,\dots, R$, we use the observed data $(y,t,x)$ to update the parameters and the augmented variables and impute the missing potential outcomes $y^{mis}$.

\vspace{0.25cm}
\noindent {\em Cluster Allocation.}
The latent variables $S_i^{(t)}$ identifies the cluster allocation for each units $i \in \{1,\dots,n\}$ at the treatment level $t$. Its posterior distribution is a multinomial distribution where
\begin{equation*}
    \mathbb{P}\{S_{i}^{(t)}=l \} \propto \omega_{l}^{(t)}(\x_i)\Norm(y_i;\eta_{l}^{(t)},\sigma_{l}^{(t)2}),
\end{equation*}
for $i=1,\dots,n$ and $l=1,\dots,L$, with $\omega^{(t)}_{l}$ defined as:
\begin{equation*}
    \omega_{l}^{(t)}(\x_i)=\Phi(\alpha_{l}^{(t)}(\x_i)) \prod_{r<l} (1-\Phi(\alpha_{l}^{(t)}(\x_i))),
\end{equation*} 
for $l=1,\dots,L-1$ and with $\Phi(\alpha_{L}^{(t)}(\x_i))=1$.

\vspace{0.25cm}
\noindent {\em Cluster Specific Parameters.}
Thanks to the latent variables $S_i^{(t)}$, that cluster the units by the value of their outcome for the treatment level $t$, we know for each cluster $l \in \{1,\dots, L\}$, the allocated units and we can update the values of the parameters from their posterior distributions:
\begin{align*}
    \eta_{1}^{(t)} & \sim  \Norm\left( V_1^{-1}\times \left(\frac{\sum_{\{i: S_{i}^{(t)}=1\}}y_i(t)}{\sigma_{1}^{(t)2}} +\frac{\mu_\eta}{\sigma^2_{\eta}} \right),V_1^{-1}\right);\\
\eta_{l}^{(t)} & \sim \Norm\left(V_l^{-1}\times \left(\frac{ \sum_{\{i: S_{i}^{(t)}=l\}}y_i(t)}{\sigma_{l}^{(t)2}} +\frac{\mu_\eta}{\sigma^2_{\eta}} \right),V_l^{-1}\right)\iv_{\{\eta_{tl}>\eta_{t(l-1)}\}} , \mbox{ for } l=2,\dots,L;\\
    \sigma_{l}^{(t)2} & \sim \mbox{InvGamma}\left(\gamma_1+ \frac{n_{l}^{(t)}}{2},\gamma_2+\frac{\sum_{\{i: S_{i}^{(t)}=l\}} (y_i(t)-\eta_{l}^{(t)})^2}{2}\right), \mbox{ for } l=1,\dots,L;
\end{align*}
where $V_l=n_{l}^{(t)}/\sigma_{l}^{(t)2} +1/\sigma^2_{\eta}$, $\iv_{\{\cdot\}}$ is the indicator variable introduced in $\eta_{l}^{(t)}$ distribution, to resolve the label switching problem in the MCMC chains, and $n_{l}^{(t)}$ is the number of units allocated in the $l$-th cluster.

\vspace{0.25cm}
\noindent {\em Augmentation Scheme.}
In order to sample from $\{\alpha_{l}^{(t)}(\x)\}_{l=1}^L$ and the corresponding weights $\{ \omega_{l}^{(t)}(\x)\}_{l=1}^L$, we need a data augmentation scheme. The idea was developed by \citet{albert2001sequential} and borrowed by \citet{rodriguez2011nonparametric} to obtain exact Bayesian inference for binary regression and computationally easy to include it in the Gibbs sampling \citep{albert2001sequential}. We can impute the augmented variables $Z_{l}^{(t)}(x_i)$ by sampling from its full conditional distribution \citep{rodriguez2011nonparametric}:
\[
    Z_{l}^{(t)}(x_i)|S_{i}^{(t)},\alpha_{l}^{(t)}(\x_i) \sim \begin{cases}
    \Norm(\alpha_{l}^{(t)}(\x_i),1)\iv_{\R^+} \mbox{ if } S_{i}^{(t)}=l,\\
    \Norm(\alpha_{l}^{(t)}(\x_i),1)\iv_{\R^-} \mbox{ if } S_{i}^{(t)}<l.
\end{cases}
\]
The mean, $\alpha_{l}^{(t)}(\x_i)$, of the previous normal distributions is obtained from:
\[
    \alpha_{l}^{(t)}(\x_i)=\phi\left(\frac{\omega_{il}^{(t)}(\x_i)}{\prod_{r<l}(1-\Phi(\x_i^T\beta_{l}^{(t)})}\right)= \phi\left(\frac{\omega_{ir}^{(t)}(\x_i)}{1-\sum_{r<l}\omega_{ir}^{(t)}(\x_i)}\right);
\]
where $\phi(\cdot)$ is the continuous density function of Gaussian distribution. 

\vspace{0.25cm}
\noindent {\em Confounder-Dependent Weights.}
To conclude the for-loop, the $\{\beta_{ql}^{(t)}\}_{q=0}^p =(\beta_{0l}^{(t)},\beta_{l}^{(t)})$, for  $l=1,\dots,\max(S_{i}^{(t)},L-1)$, are updated for the posterior distribution:
\begin{align*}
    \beta_{0l}^{(t)} & \sim \Norm( (1/\sigma^2_{\beta}+n)^{-1}\times (\mu_\beta/\sigma_\beta^2 +1_n^T\Tilde{\bf Z}),(1/\sigma^2_{\beta}+n)^{-1}); \\
    \beta_{l}^{(t)} & \sim \Norm_{p}( W^{-1}\times (\mu_\beta/\sigma_\beta^2 +(\Tilde{\X})^T\Tilde{\bf Z}),W^{-1});
\end{align*}
where $1_n$ is a $n$ vector of ones, $W=I_{p}/\sigma^2_{\beta}+(\Tilde{\X})^T\Tilde{\X}$, $I_{p}$ is a  $p\times p$ diagonal matrix, $\Tilde{\X}$ is a matrix such that it is composed by the rows $i$ in $\X$, such that $S_{i}^{(t)}\leq l$, and $\Tilde{\bf Z}$ is a vector composed by the $z_{l}^{(t)}(x_i)$ for the units $i$ such that $S_{i}^{(t)}\leq l$.

{\vspace{0.8cm}
\spacingset{1}
\begin{algorithm}
\caption{Estimation Confounder-Dependent Mixture Model (CDBMM)}\label{alg:gibbs}
\vspace{0.15cm}
{\bf Inputs:} the observed data $(y,t,x)$.

{\bf Outputs:} 

\quad - posterior distributions of parameters: $\eta$, $\sigma$, and $\beta$;

\quad -  posterior distribution over the space of partitions of the units.
\vspace{0.05cm}

{\bf Procedure:}
\begin{algorithmic}
\State Initialization of all parameters and latent variables.
\For{$r \in \{1,\dots,R\}$}
\For{$t\in \{0,1\}$}
    \State Compute $\omega^{(t)}(x_i)$ for $i=1,\dots,n$;
    \State Draw $S_{i}^{(t)}$ for $i=1,\dots,n$;
    \State Draw $\eta^{(t)}$ and  $\sigma^{(t)}$;
    \State Compute $\alpha^{(t)}(\x_i)$ for $i=1,\dots,n$;
    \State Draw $z^{(t)}(\x_i)$ for $i=1,\dots,n$;
    \State Draw $\beta^{(t)}$.
\EndFor
\EndFor
\end{algorithmic}
\end{algorithm}}

\section{Simulated scenarios}
In each scenario reported in Section 3, we assume that the treatment variable and the confounders are binary and, for $i =1, \dots, n$ we simulate them from $X_{1i} \sim \mbox{Be}(0.4)$, $X_{2i} \sim \mbox{Be}(0.6)$, $X_{3i} \sim \mbox{Be}(0.3)$, $X_{4i} \sim \mbox{Be}(0.5)$, $X_{5i} \sim \mbox{Be}(0.2)$ and  $T_i \sim \mbox{Be}(\mbox{logit}(0.4X_{1i}+0.6X_{2i}))$, for all scenarios except for Scenario 5 where the treatment is defined as $T_i \sim \mbox{Be}(\mbox{logit}(0.4X_{1i}+0.6X_{2i}-0.3X_{3i}+0.2X_{4i}X_{5i}))$, where $\mbox{Be}(\theta)$ represents a Bernoulli random variable with success probability $\theta$.

Each scenario assumes a different conformation of the groups. Each group is obtained by introducing, for each unit, categorical variables $G_i$ with $k$ categories and allocating them according to covariates values.
Note that we use the words \emph{group} and \emph{cluster} consistently with the previous section. Conditionally on $G_i=g$ we simulate both  potential outcomes as $Y_i(0)|G_i=g \sim \Norm(\eta_{g}^{(0)},\sigma^{(0)}_{g})$ under control and $Y(1)|G_i=g\sim \Norm(\eta^{(1)}_{g},\sigma^{(1)}_{g})$ under treatment.

In Scenario 1, we investigate a situation in which the expected value of the outcome decreases with the treatment, but the intensity of this decrease varies across different groups. We set ${\eta}^{(0)}=(2,4,6)$ and ${\eta}^{(1)}=(0,3,6)$, and assume that the variance within each group is constant, with $\sigma_{g}^{(0)}=\sigma_{g}^{(1)}=0.3$ for $g=1,2,3$. This results in GATEs of $(-2,-1,0)$ respectively in the three groups.
The units are allocated in the three groups according to the values of the covariates: $g=1$ when $X_{1i}=X_{2i}=0$, $g=2$ when $X_{1i}=1$, and $g=3$ when $X_{1i}=0$ and $X_{2i}=1$.

Scenario 2 examines a setting where the population has different outcomes under the control group but similar outcomes under treatment. To achieve this, we set ${\eta}^{(0)}=(0,2.2,4.4)^T$ and $\eta_{g}^{(1)}=0$ for each $g\in{1,2,3}$. Like Scenario 1, we assume that $\sigma_{g}^{(0)}=\sigma_{g}^{(1)}=0.2$, resulting in GATEs of $(0,-2.2,-4.4)$ for each group. The group allocation is the same as Scenario 1.

In Scenario 3, we focus on a case where the groups are less separated, with the location parameters $\eta^{(0)}$ and $\eta^{(1)}$ being closer to each other and different variances between groups and for treatment and control. Specifically, we set $\eta^{(0)}=(1,2,3)$, $\eta^{(1)}=(0,1.5,3)$, $\sigma^{(0)}=(0.2,0.25,0.25)$, and $ \sigma^{(0)} =(0.25,0.3,0.2)$. This results in GATEs of $(-1,-0.5,0)$. The group allocation is the same as Scenario 1.

Scenario 4 considers a situation in which there are four groups, combining features from the previous scenarios. Specifically, we consider the values within $\eta^{(0)}$ and $\eta^{(1)}$ to be close to each other, as in Scenario 3, and assume a different behavior under treatment and control in terms of heterogeneity as in Scenario 2. Specifically, the outcomes under treatment show four different clusters, while the outcomes under control underline three different clusters. This is achieved by setting $\eta^{(0)}=(1,2,3,3)$, $\eta^{(1)}=(0,1.5,3,4.5)$, and $\sigma_{g}^{(0)}=\sigma_{g}^{(1)}=0.2$ for $g=1,2,3,4$. The GATE results to be $(-1,-0.5,0,1.5)$ respectively in the four groups. The units are allocated in the four groups according with the covariates values: $g=1$ when $X_{1i}=0$ and $X_{2i}=1$, $g=2$ when $X_{1i}=X_{2i}=0$, $g=3$ when $X_{1i}=X_{2i}=1$, and $g=4$ when $X_{1i}=1$ and $X_{2i}=0$.

In Scenario 5, We consider all the five confounders $\X_1,\dots, \X_5$ as well as characterization of the groups. Moreover, in this scenario we also increase the number of groups up to 5, describing a more complex and heterogeneous setting. The cluster-spesific parameters are set to $\eta^{(0)}=(2,2,3,4.5,6.5)$, $\eta^{(1)}=(0,1,2.5,5,7.5)$, $\sigma_{g}^{(0)}=\sigma_{g}^{(1)}=0.2$ for $g=1,2,3,4,5$. We obtain GATEs equal to $(-2, -1, -0.5, 0.5, 1)$. The units are allocated in the five groups according to the values of the five covariates: $g=1$ when $X_{1i}=X_{2i}=1$, $g=2$ when $X_{1i}=0$ and $X_{3i}=1$, $g=3$ when $X_{1i}=X_{3i}=0$ and $X_{4i}=1$, $g=4$ when $X_{1i}=X_{3i}=X_{4i}=0$, and $g=5$ otherwise.

Scenario 6 is similar to Scenario 1 and Scenario 3, but with groups that are even closer and with bigger variance, such that the marginal distributions for the treated and control outcomes, respectively, are not multimodal. In particular, we have three groups with $\eta^{(0)}=(1.5,2,2.5)$, $\eta^{(1)}=(1,1.75,2.5)$, and $\sigma_{g}^{(0)}=\sigma_{g}^{(1)}=0.3$ for $g=1,2,3$. We obtain GATEs equal to $(-0.5, -0.25, 0)$. The group allocation is the same as Scenario 1.

In scenario 7, we study the degenerative case of heterogeneity, such that we have only one group, with $\eta^{(0)}=2$, $\eta^{(0)}=3$, and $\sigma_{g}^{(0)}=\sigma_{g}^{(1)}=0.5$. The average treatment effect is equal to 1.

In each setting, the sample size is fixed to $n=500$. For each scenario, we simulate $100$ samples. We set the number of groups as following: $k=3$ for Scenarios 1-2-3-6, $k=4$ for Scenario 4, $k=5$ for scenario 5, and $k=1$ for Scenario 7.

\begin{figure}[!htp]
\begin{center}
\includegraphics[width=2in]{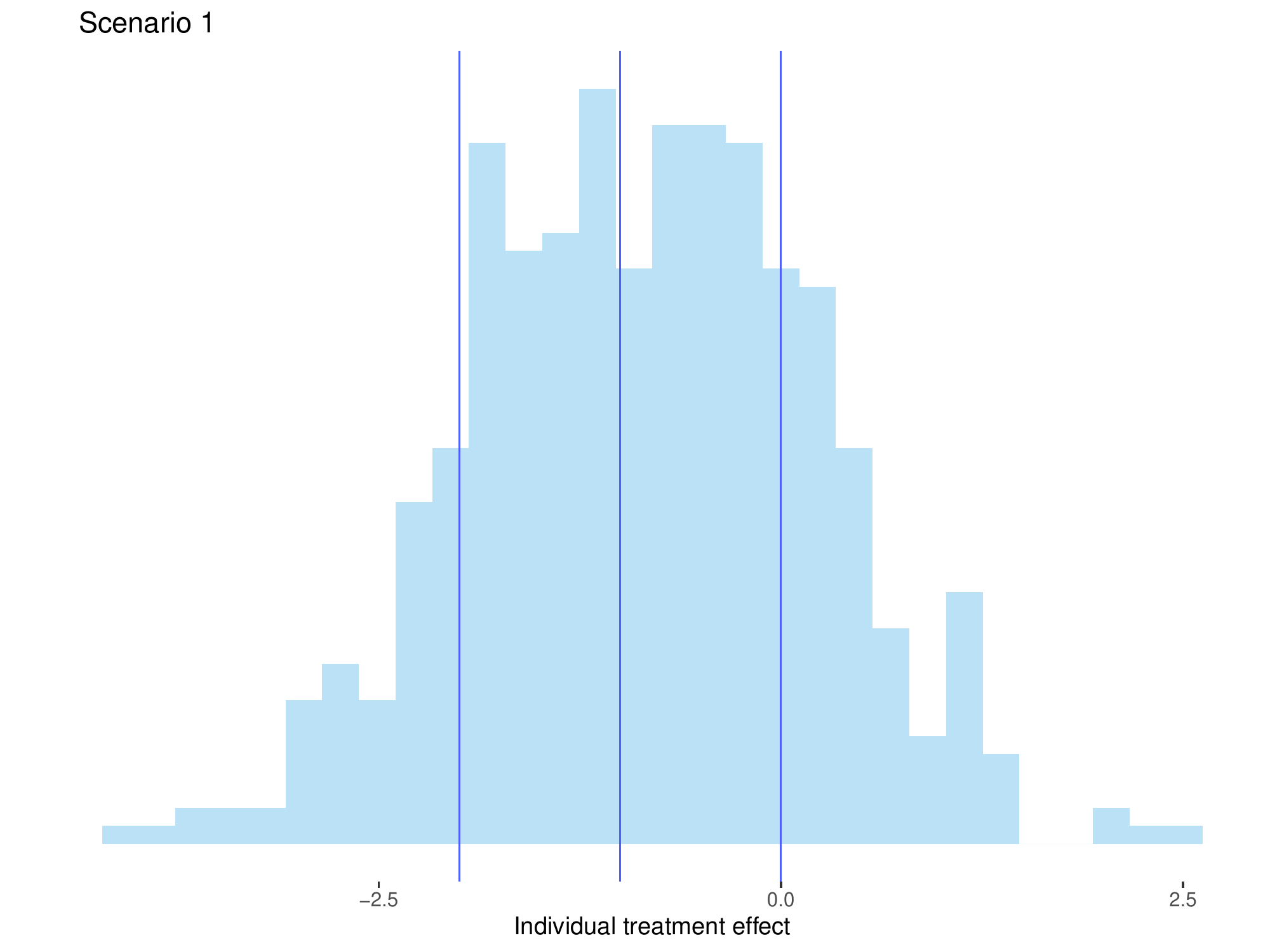}\; \includegraphics[width=2in]{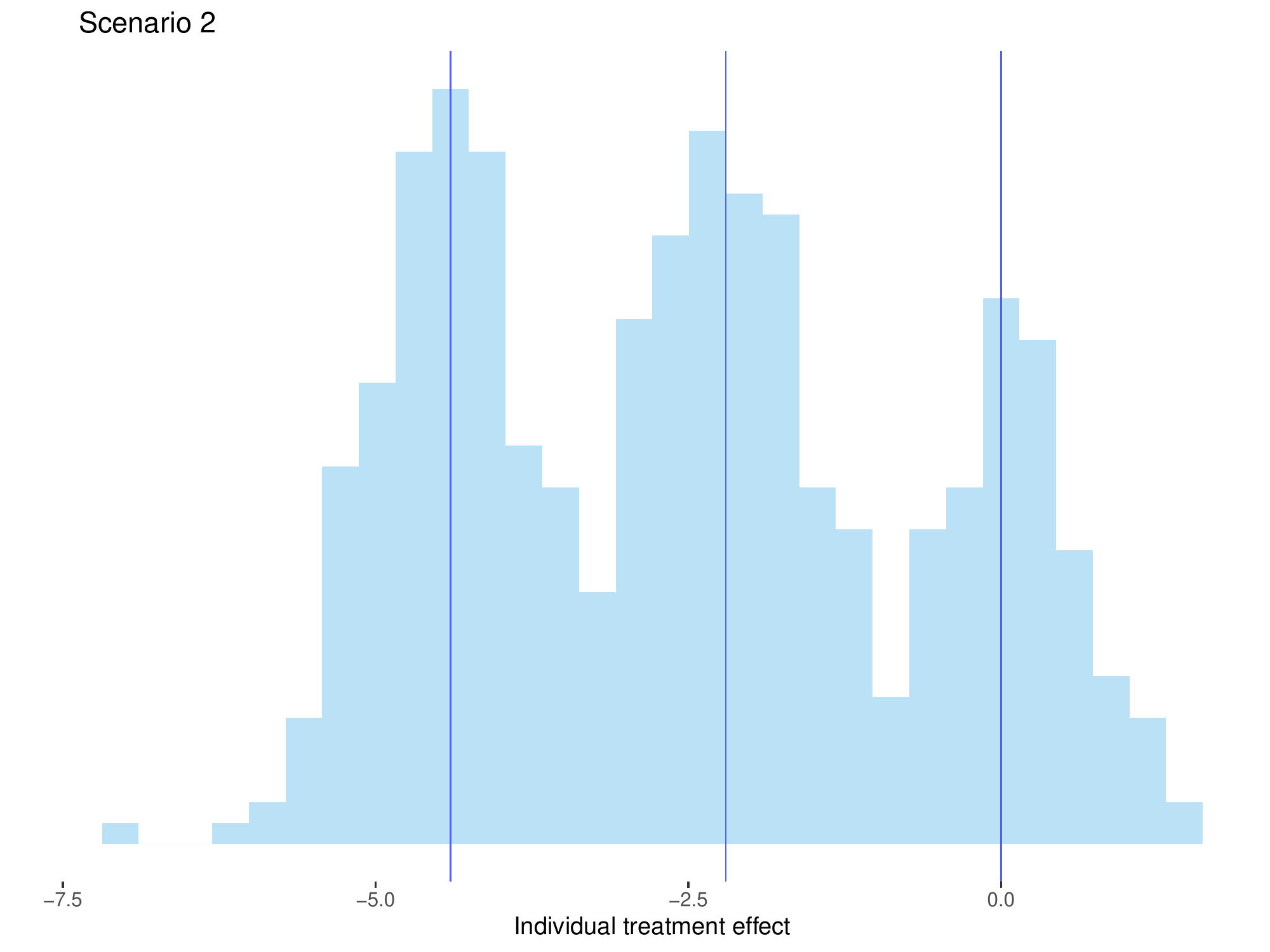}\\
\vspace{0.5em}
\includegraphics[width=2in]{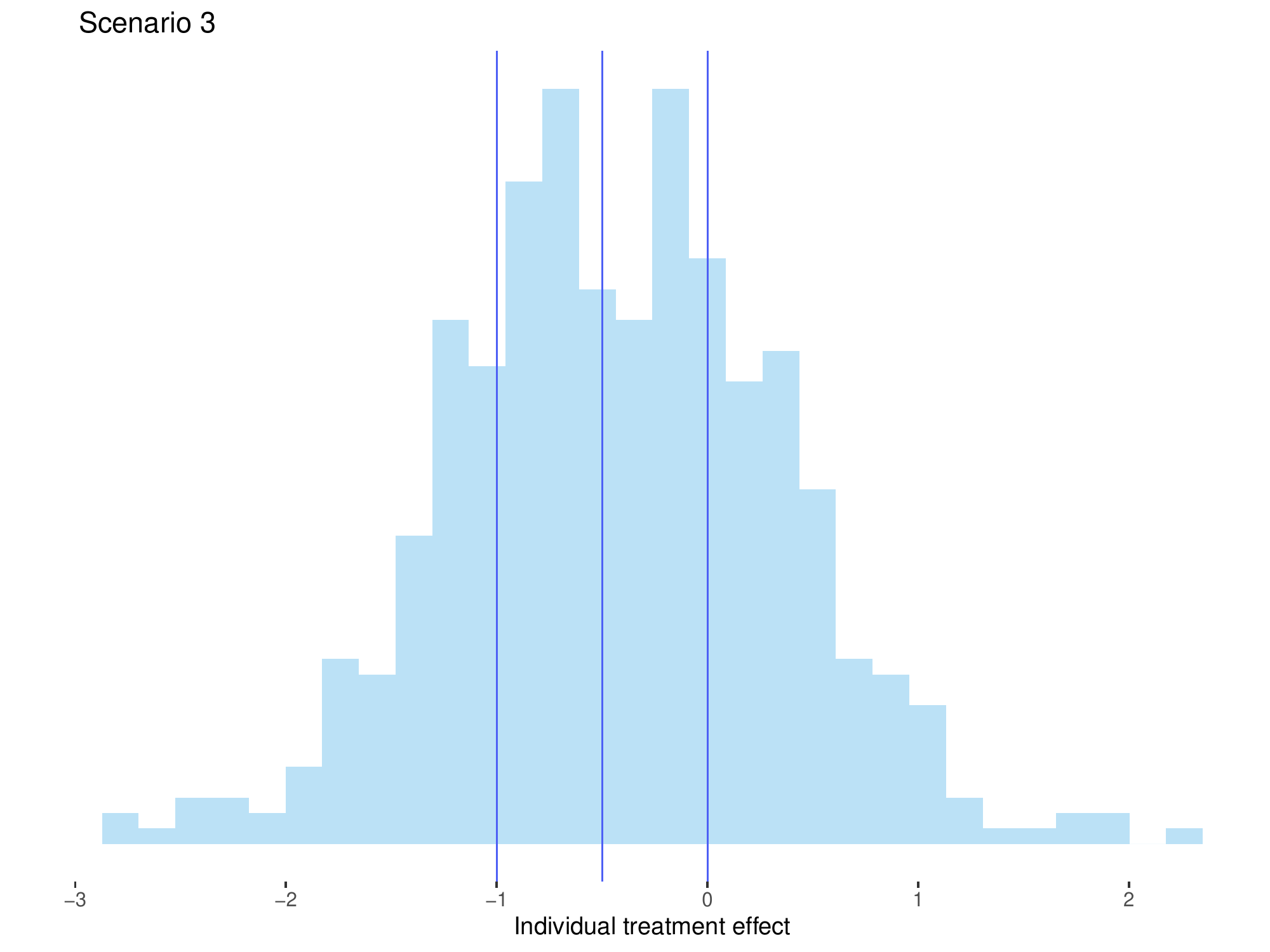}\; \includegraphics[width=2in]{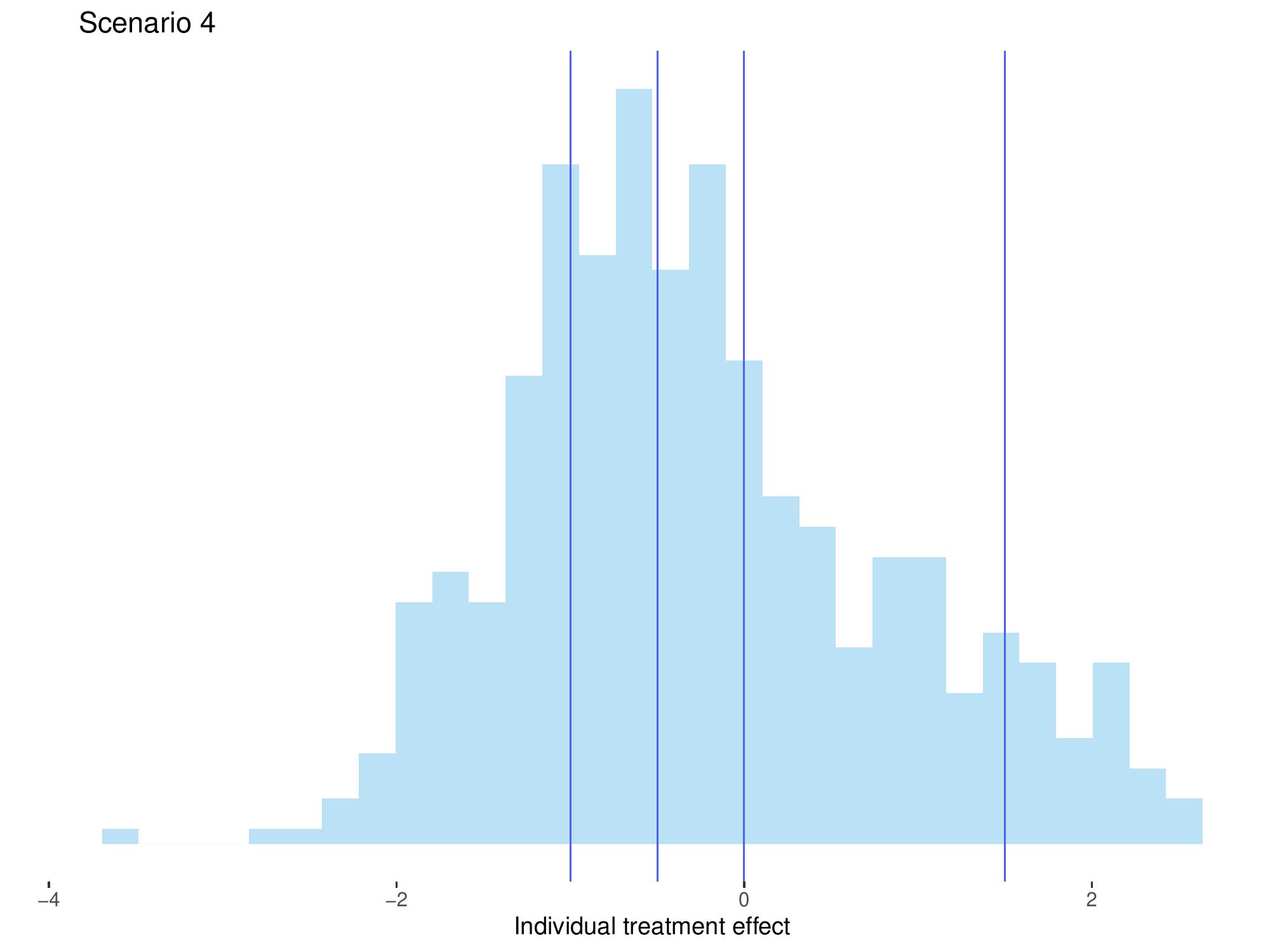}\; \includegraphics[width=2in]{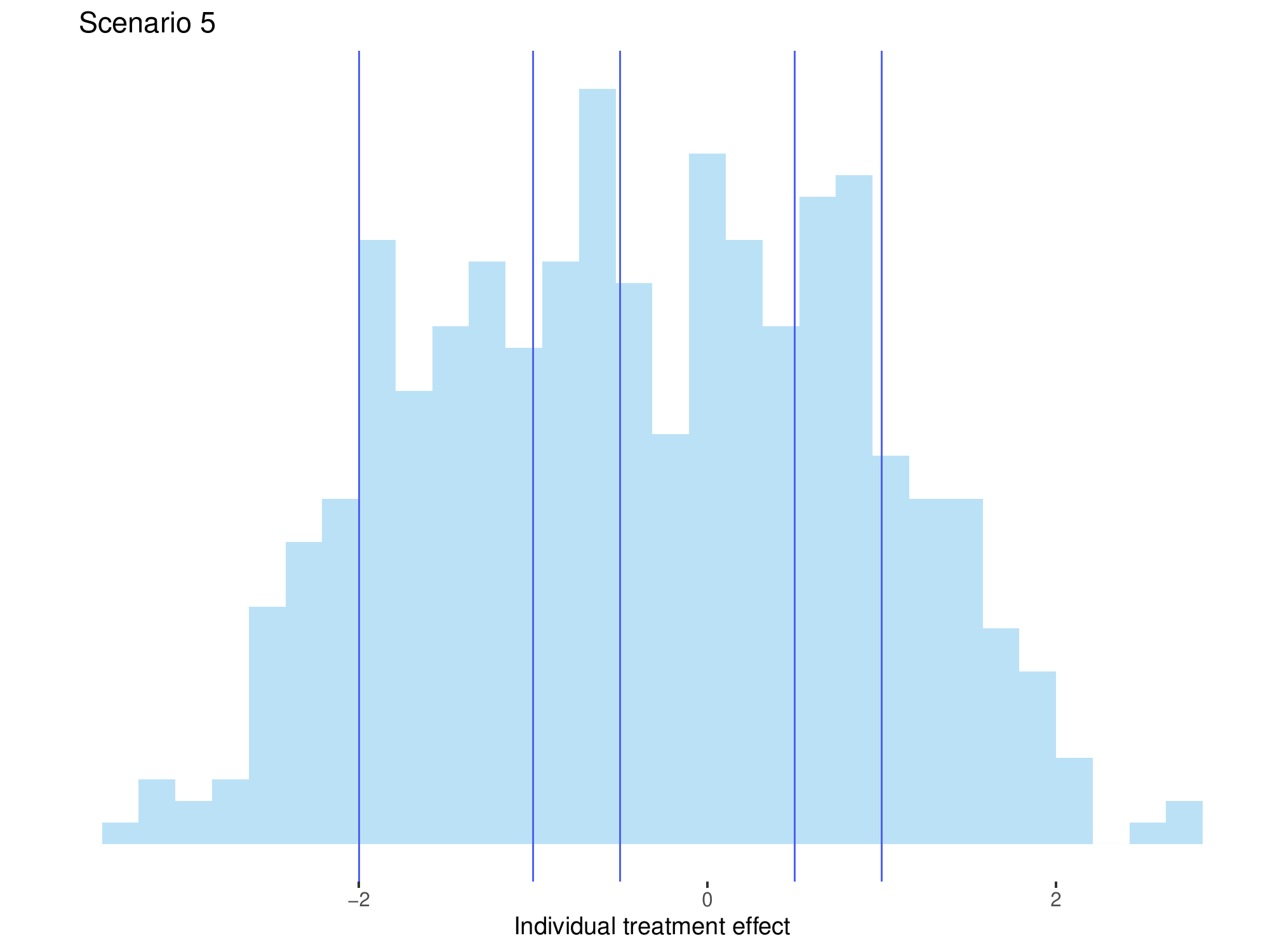}\\
\vspace{0.5em}
\includegraphics[width=2in]{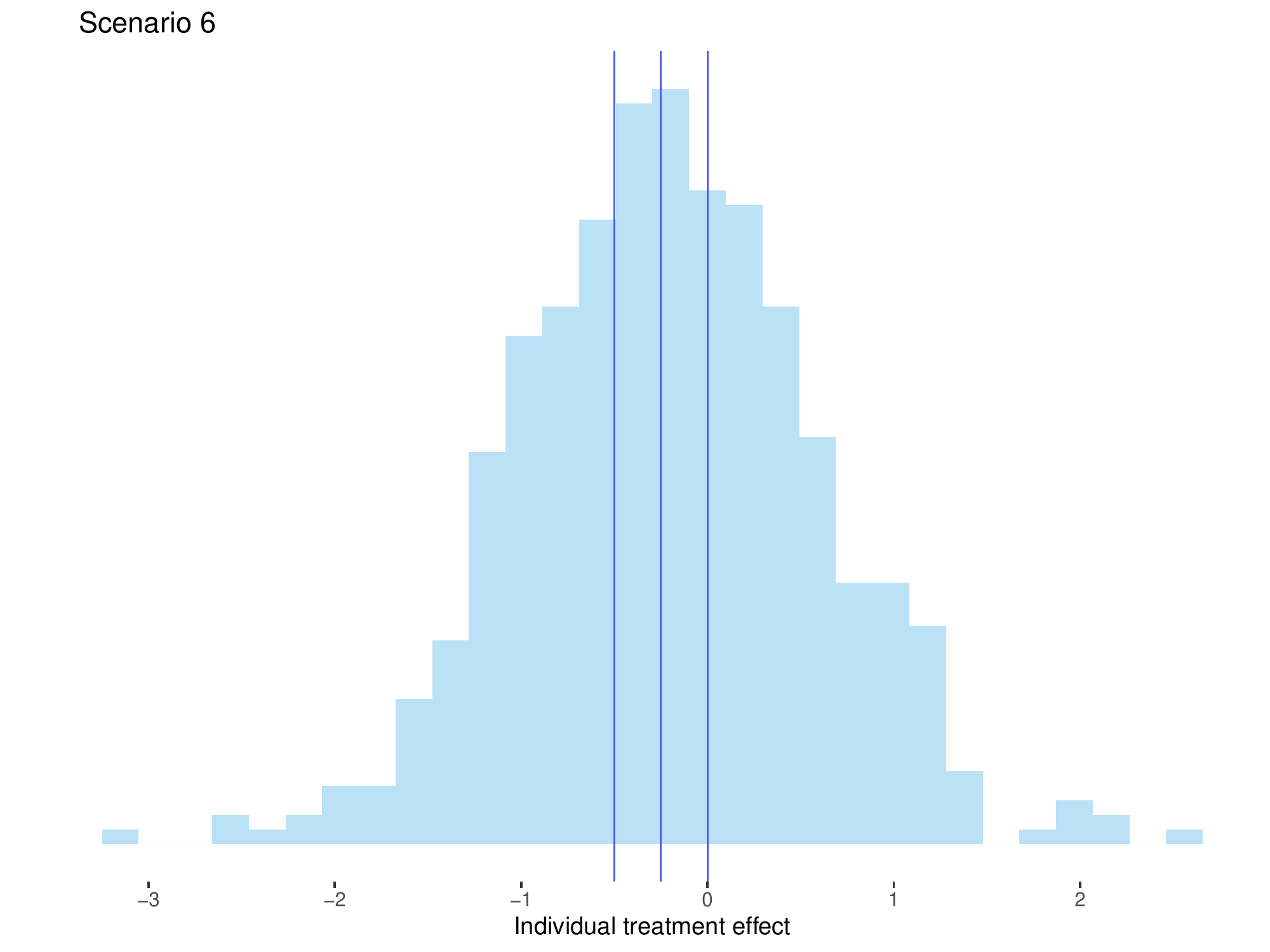}\; \includegraphics[width=2in]{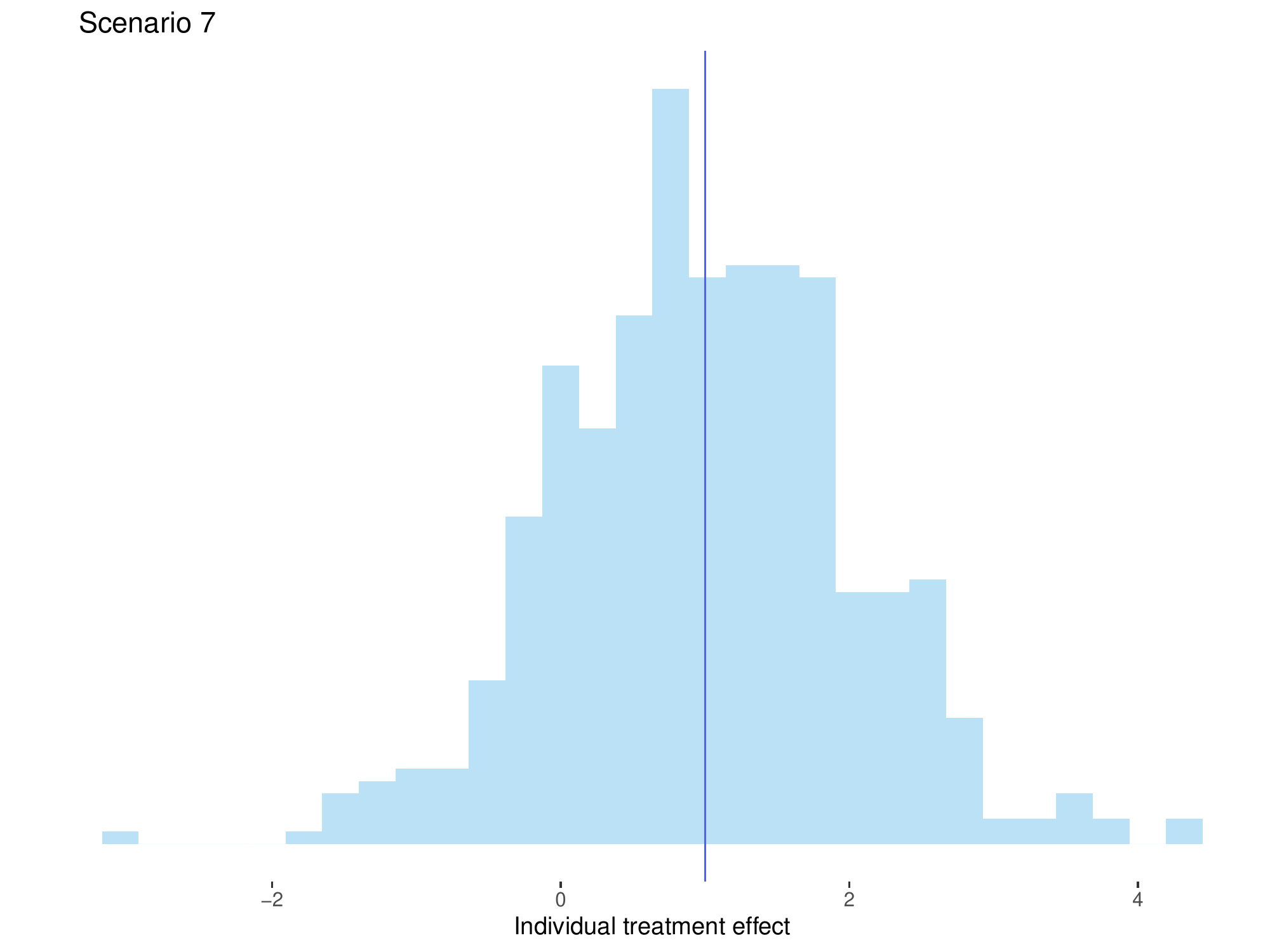}\\
\end{center}
\caption{Histograms of the distribution of the simulated causal effects in the seven scenarios. The blue lines show the values of the GATEs.\label{fig:sim_iate}}
\end{figure}

The Figure \ref{fig:sim_iate} reports the simulated distribution of the causal effects of the seven simulated scenarios. The blue vertical lines show the values of the GATE for the groups of each scenario.

For the estimation of CDBMM, we choose the same hyperparameters for each setting and such that the prior is non-informative. For the regression parameters in 
(8) and for the parameters $\eta_l^{(t)}$ and $\sigma_l^{(t)}$ in 
(9) we use the following conjugate priors
\[
\beta_{ql}^{(t)} \sim \Norm(0,20), \; \eta_l^{(t)} \sim \Norm(0,10), \mbox{ and } \sigma_l^{(t)} \sim \mbox{InvGamma}(5,1).
\]
for $t\in \{0,1\}$, $l\in\{1,\dots,20\}$---see the Supplementary Material for the choice of finite truncation of the DPSB process---, and $q$ according to the covariates considered in different settings.

\section{Additional simulation study results}
\label{sec:app_carr}
The proposed method not only produces accurate ATE estimates but also excels in estimating the GATEs. The following boxplots report the estimated GATEs for each group in the seven simulated scenarios. The medians of each GATE are in close agreement with the true values, as evident from the close alignment with the horizontal dashed lines, demonstrating the model's capability to distinguish between different groups and estimate the corresponding GATEs with high precision.

\begin{figure}[!htp]
\begin{center}
\includegraphics[width=2.3in]{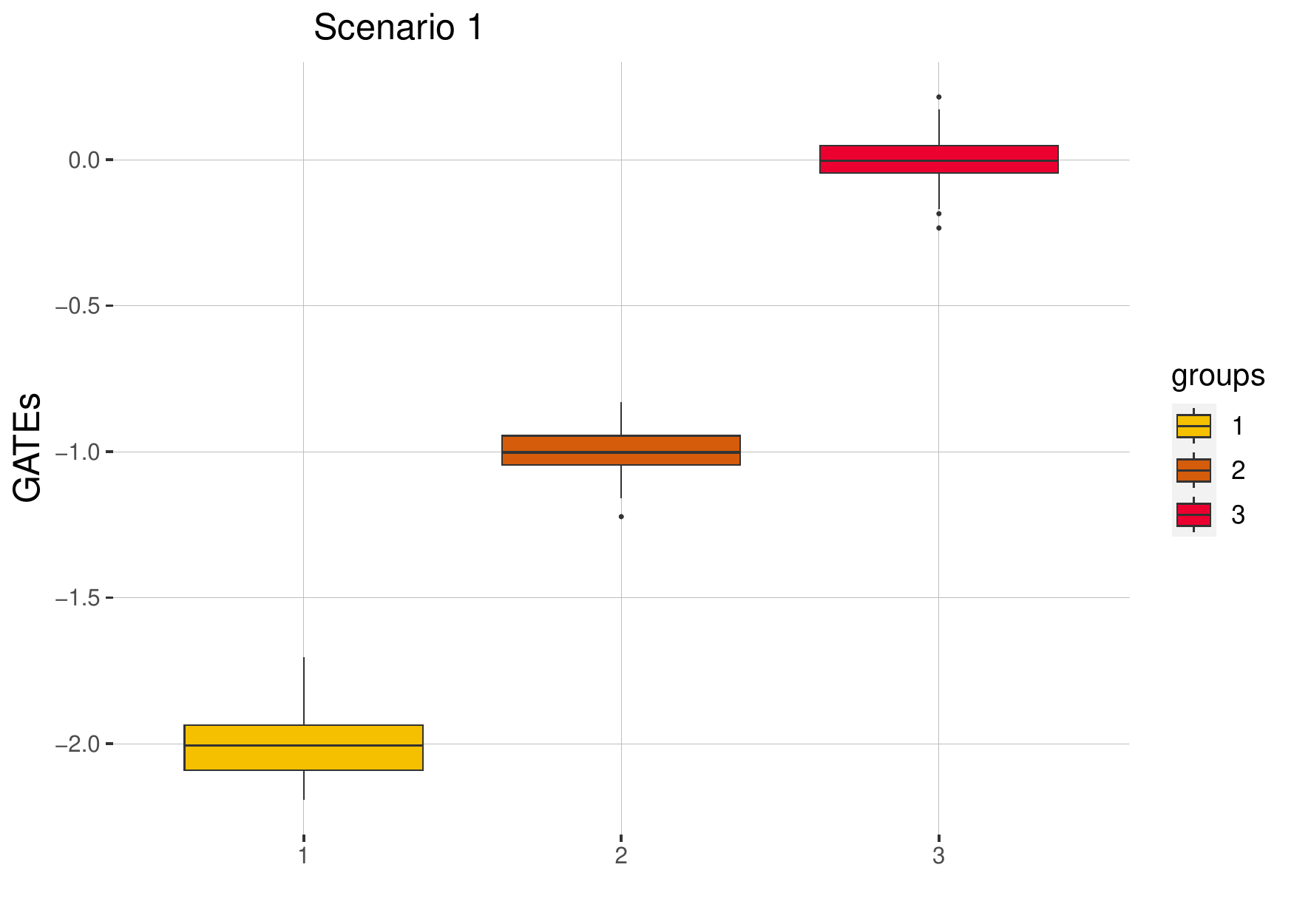}\; \includegraphics[width=2.3in]{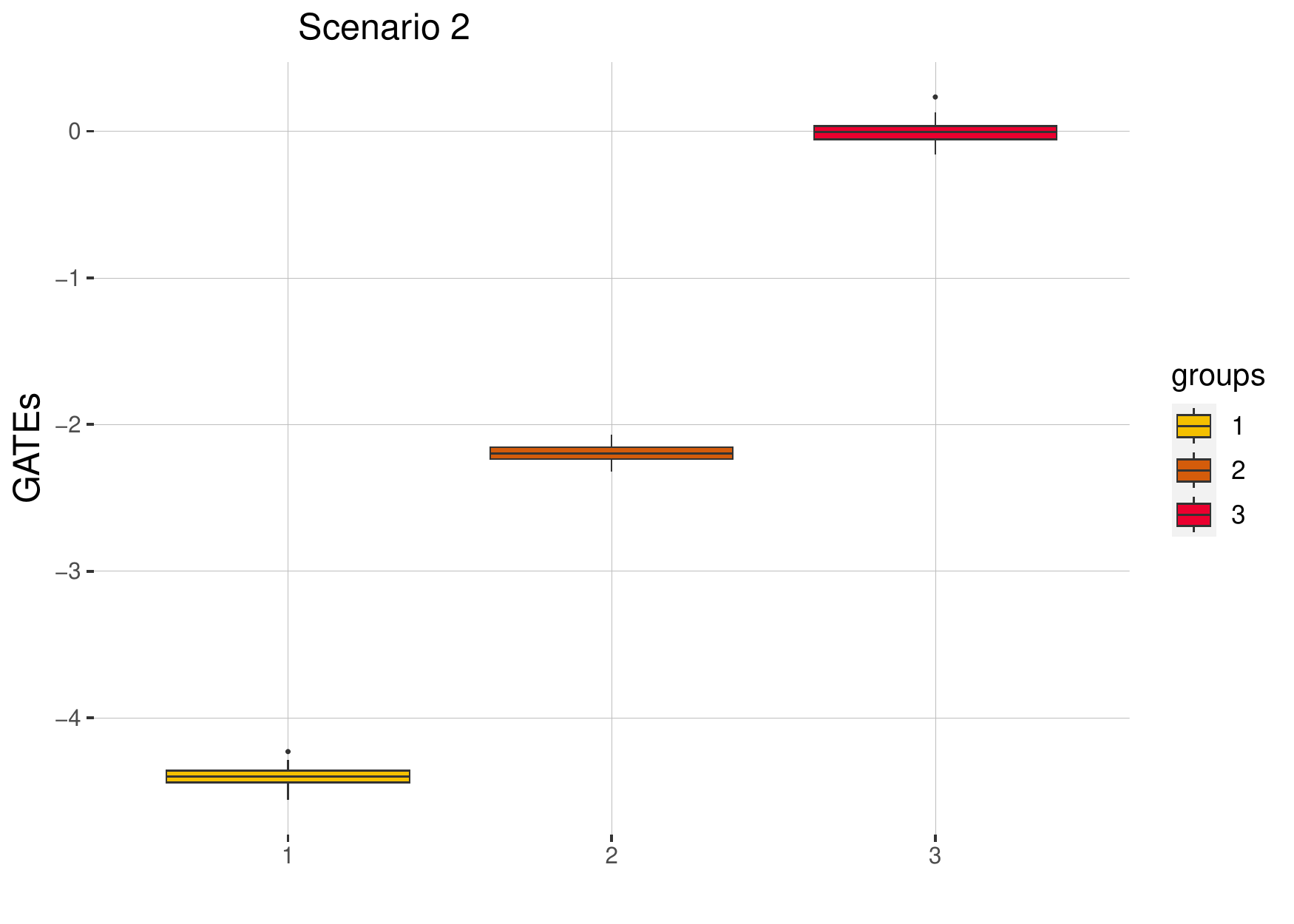}\\
\vspace{0.5em}
\includegraphics[width=2.1in]{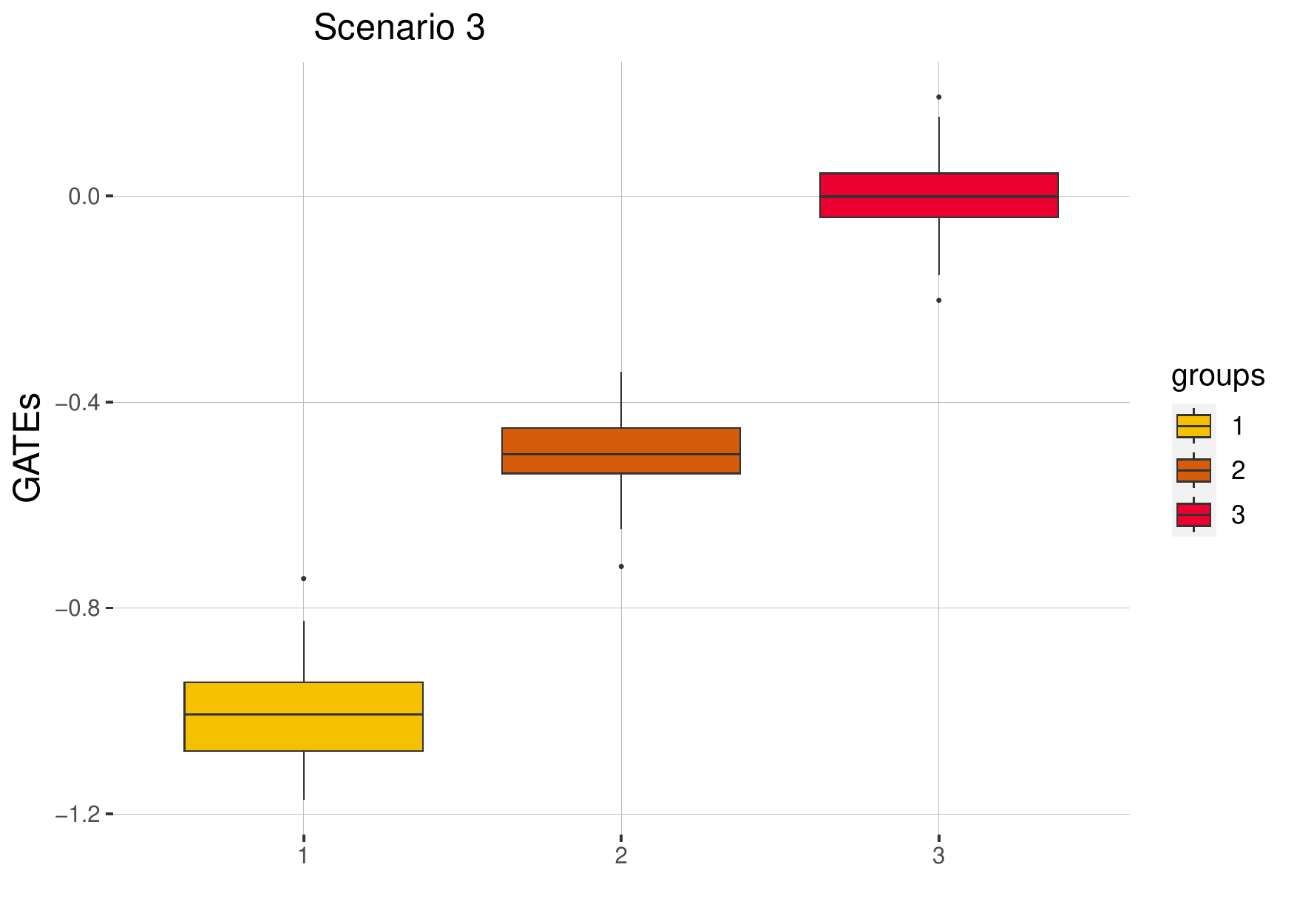}\; \includegraphics[width=2.1in]{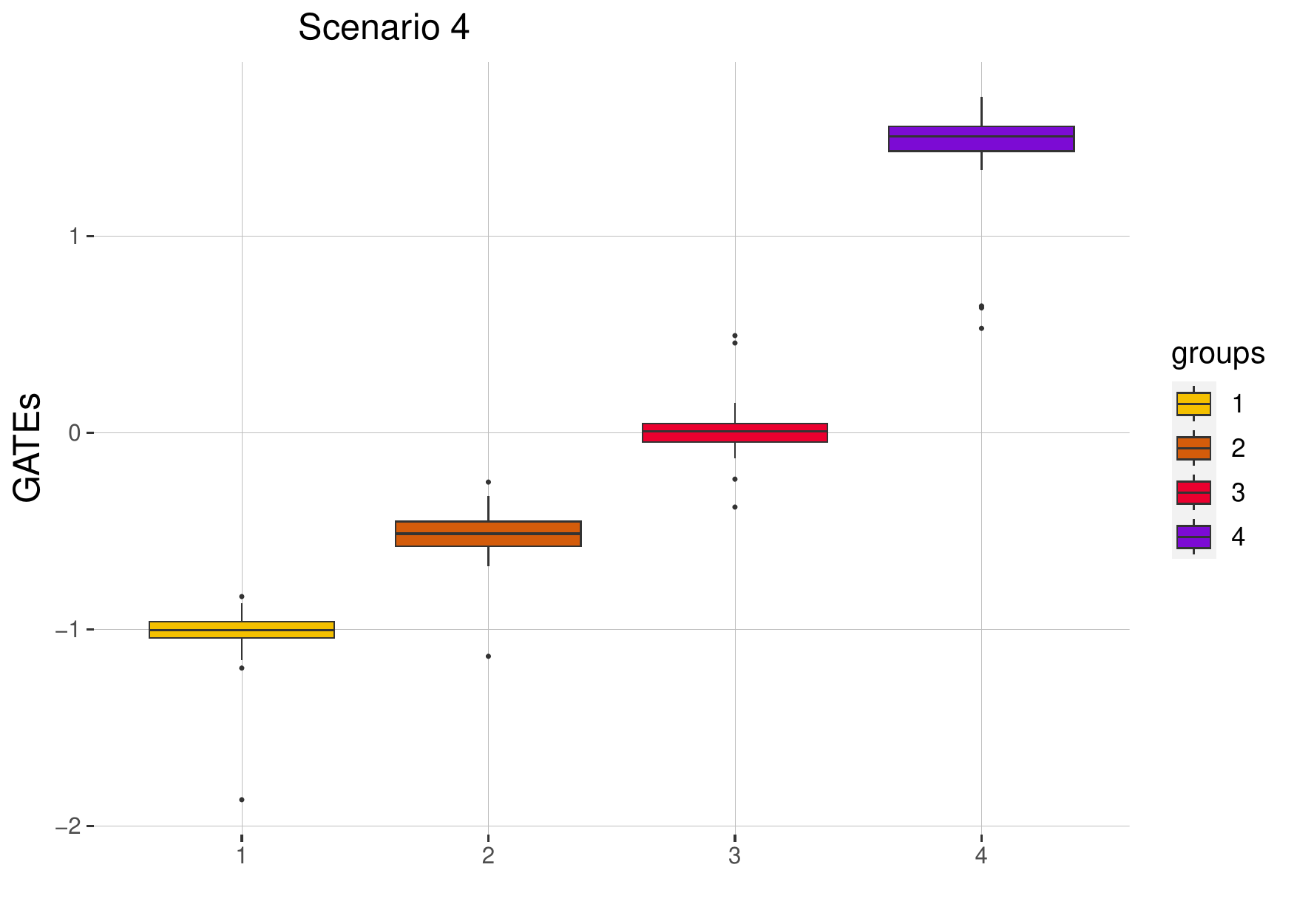}\;
\includegraphics[width=2.1in]{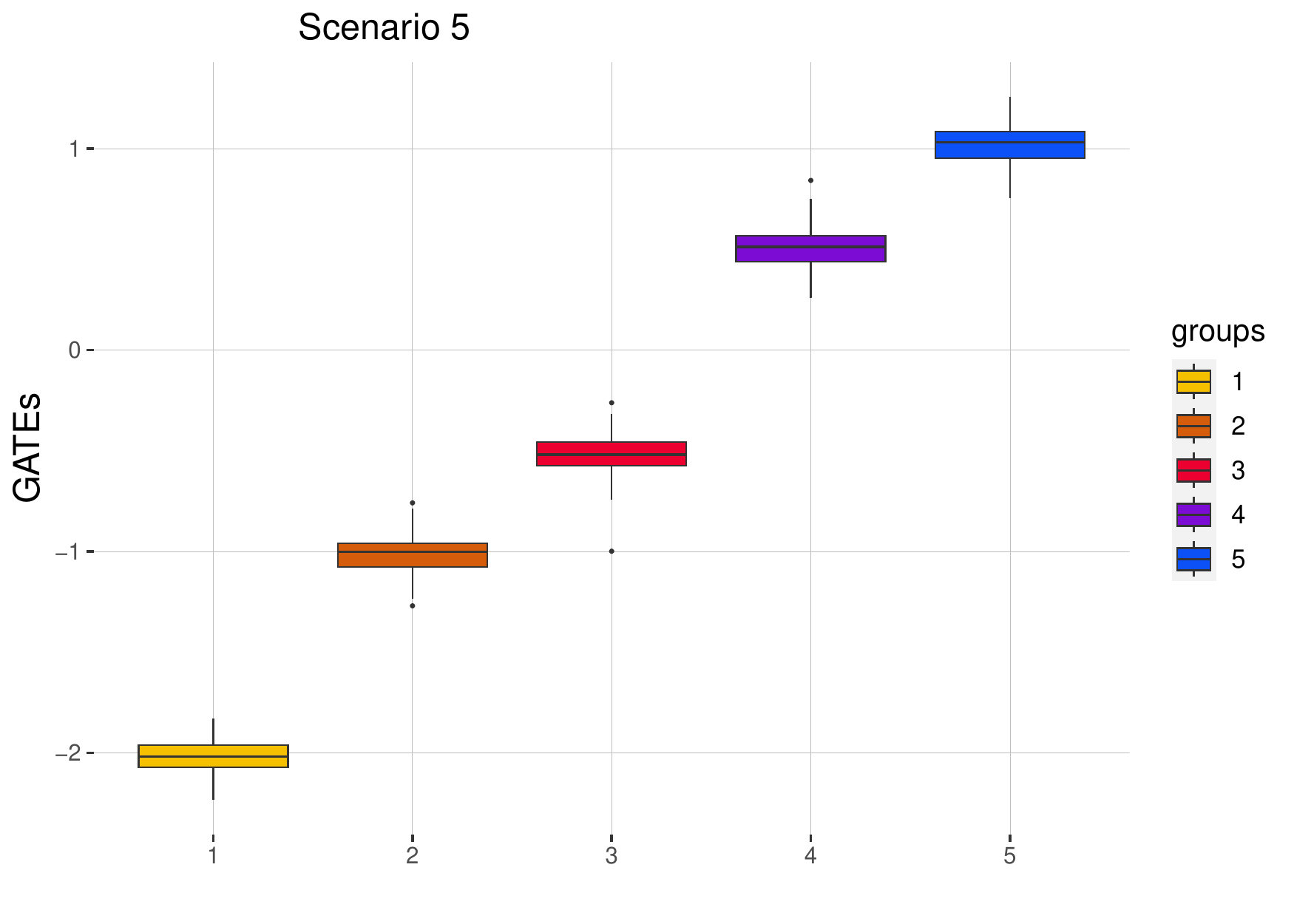}\\ 
\vspace{0.5em}\includegraphics[width=2.2in]{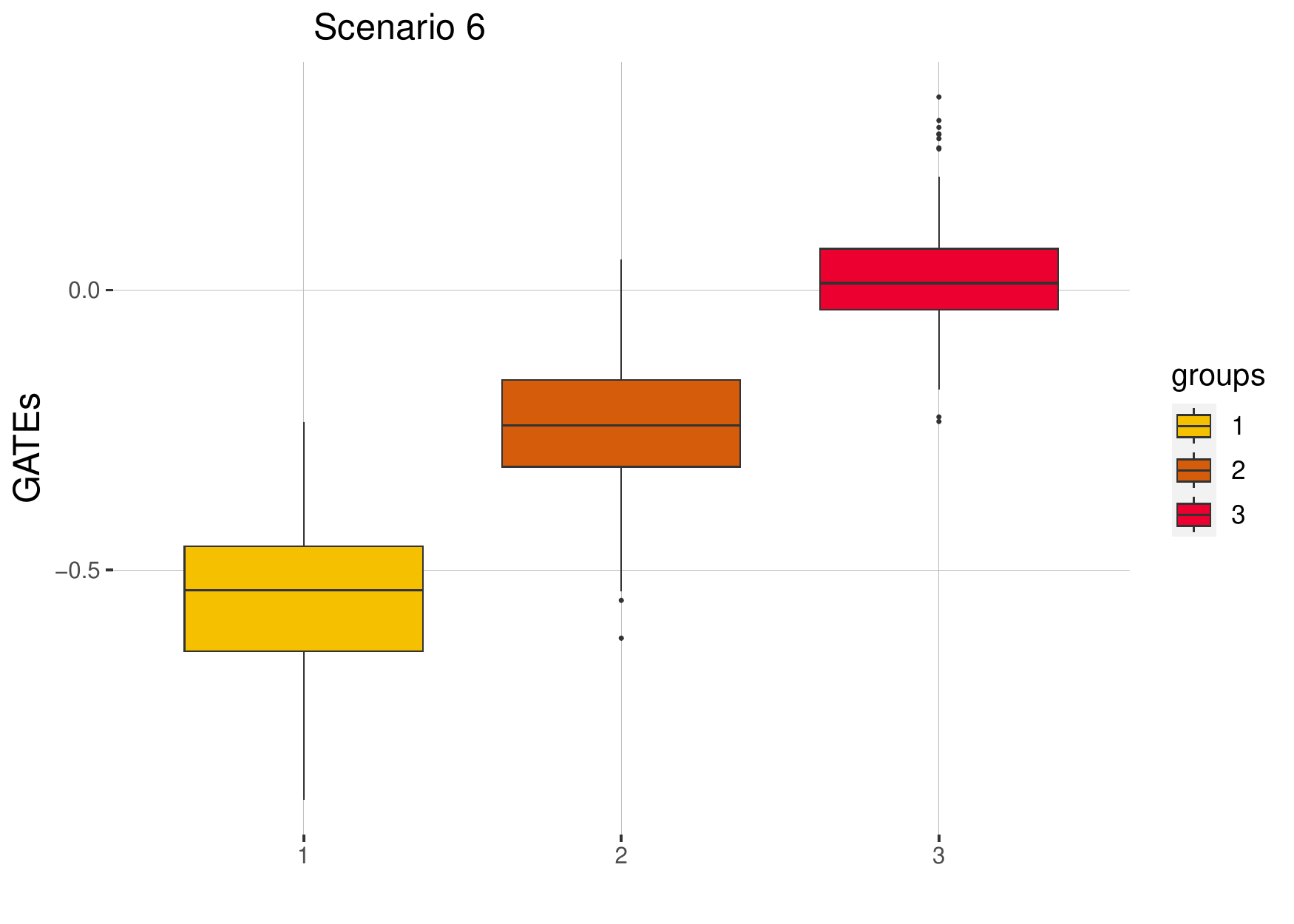}\;
\vspace{0.5em}
\includegraphics[width=2.2in]{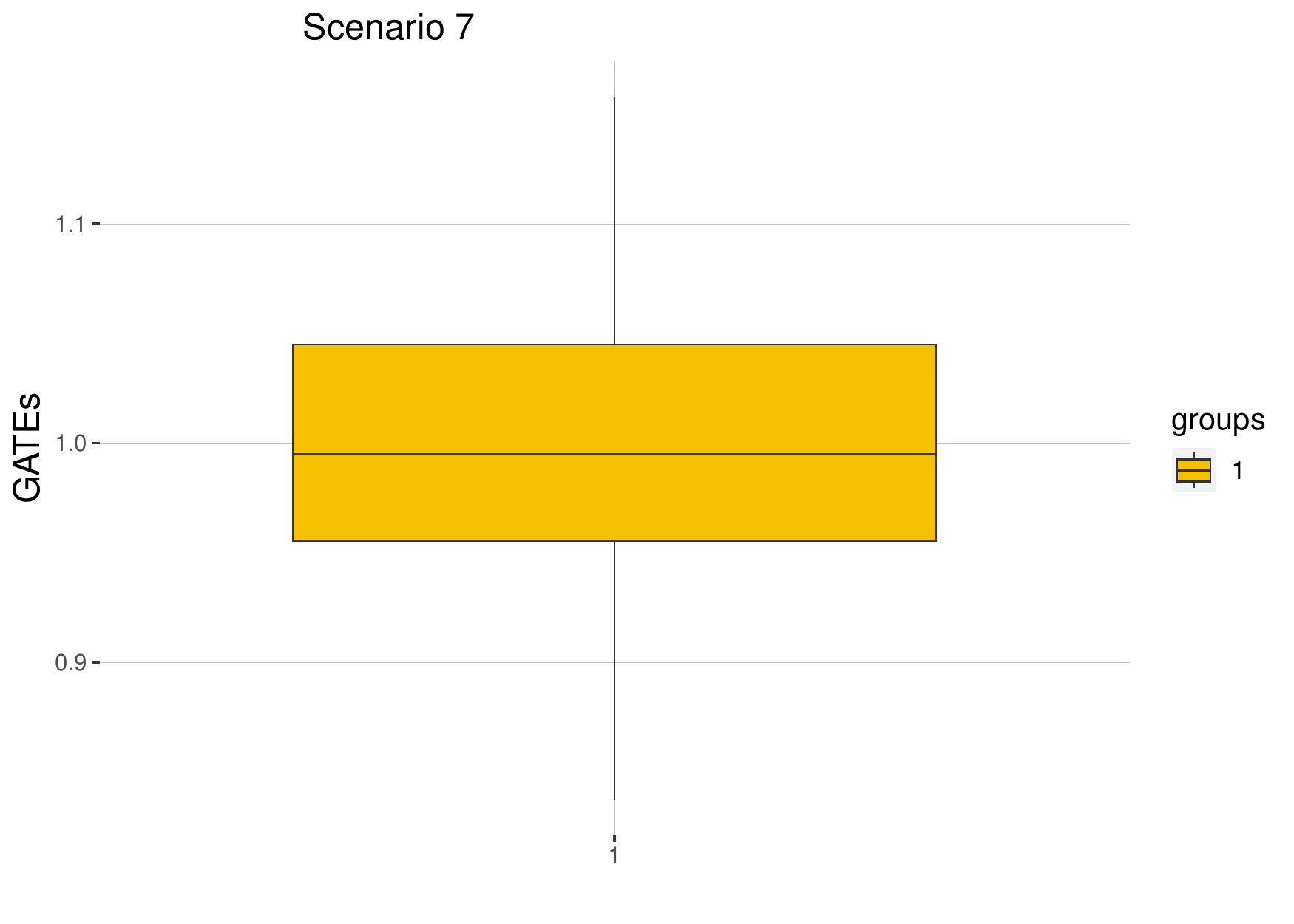}
\end{center}
\caption{GATEs for the groups in the seven simulated scenarios. The dashed lines show the true values.\label{fig:sim_cluster}}
\end{figure}

\section{Sensitivity analysis}
\label{sec:app_sens_analysis}

Using a plain Dirichlet process, the total mass parameter determines the expected prior number of groups in the data: a prior influence that frequently carries over to the posterior distribution. The BNP literature addresses this issue by focusing on two solutions: i) perform sensitivity analyses checking the results obtained with different values of this parameter and ii) change the prior structure by resorting to the more flexible Pitman-Yor process or simply placing a hyperprior on the total mass parameter. All that said, in our settings, we do not have a total mass parameter, as our stick-breaking construction (which is directly connected to the number of groups) is that of a probit stick-breaking process where the weights depend, in a probit regression framework, by regression parameters  $\beta_{ql}^{(t)}$ which, in turn, have prior distribution 
\[
\beta_{ql}^{(t)} \sim \mathcal{N}(0, \sigma^2_\beta),
\]
for $t=0,1$, $l\geq 1$,and  $q=0, 1, \dots,p$.
Despite we are already in case ii), i.e. where the stick-breaking process does not depend on a prior parameter but on a parameter that is assigned an additional layer of hyperpriors, we studied the effect of the scale parameter $\sigma^2_\beta$ repeating the experiments for $\sigma^2_\beta \in (1,20,100)$.

The results, reported in Figure \ref{fig:SA_boxplot_r1} and Table \ref{table:ARI_SA_r1} below show no significative difference among the settings.

\begin{figure}[!htb]
\begin{center}
\includegraphics[width=5in]{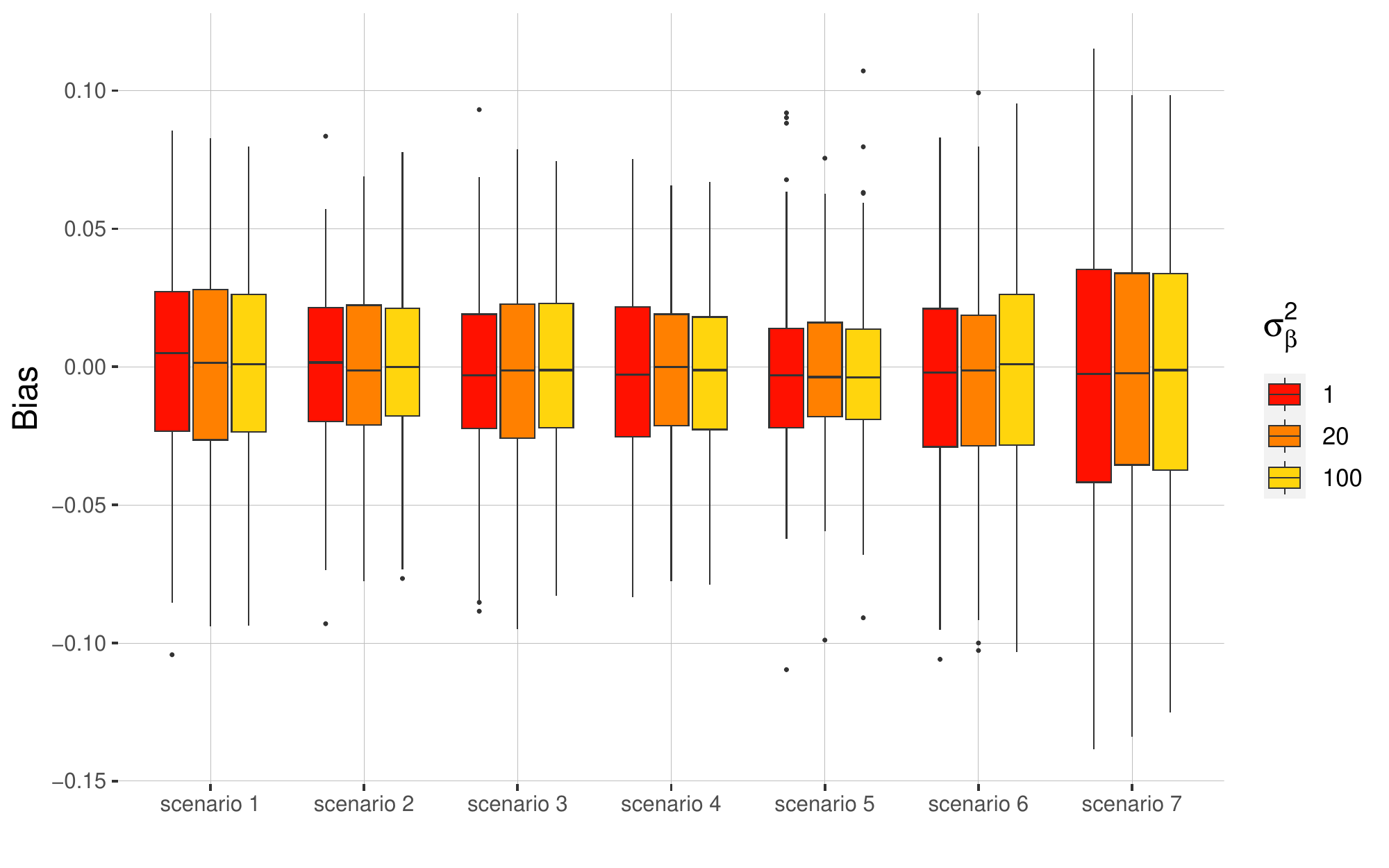}\\
\includegraphics[width=5in]{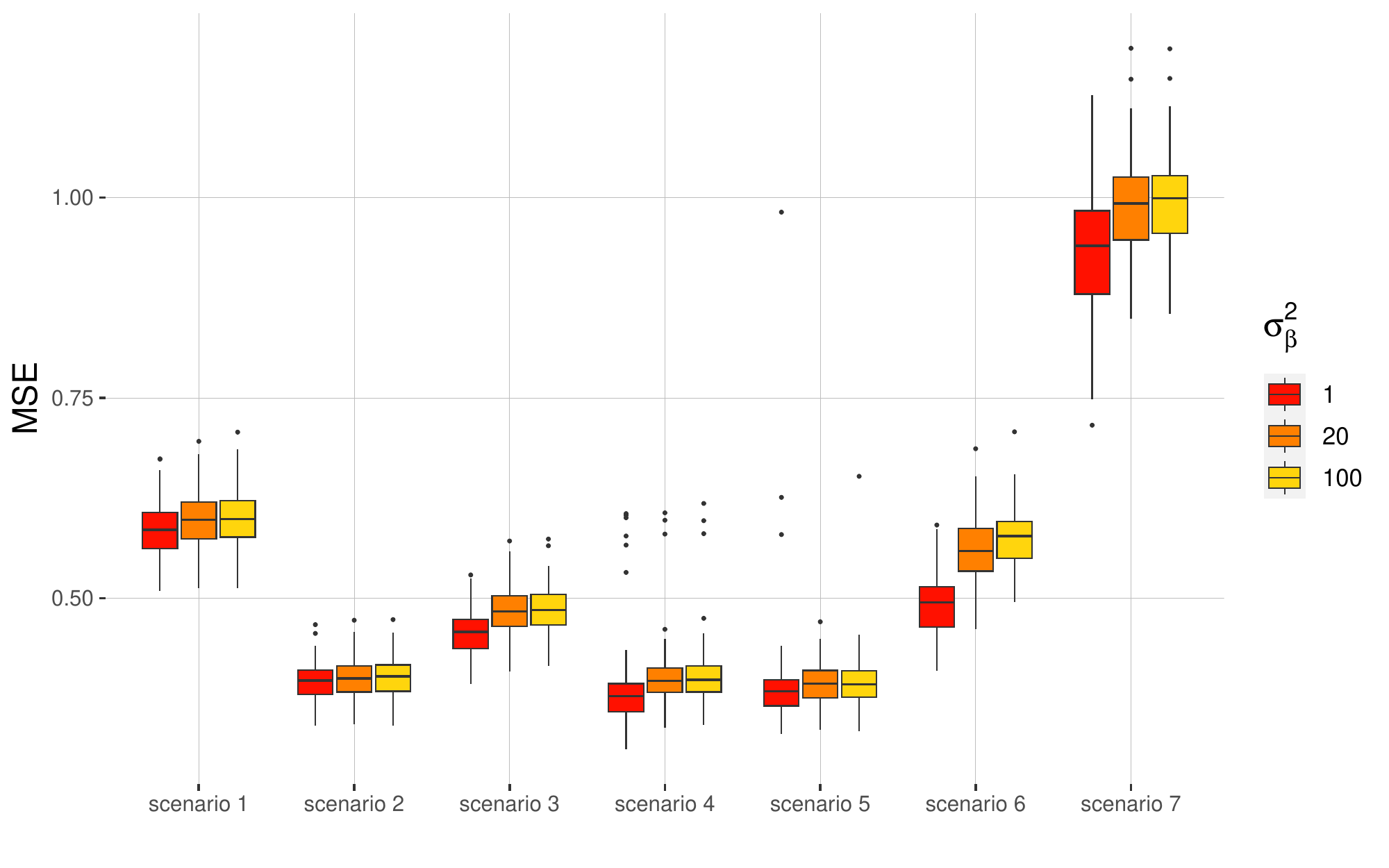}
\end{center}
\caption{Comparison of estimation of ATE in settings of sensitivity analysis: bias and mean square error (MSE). \label{fig:SA_boxplot_r1}}
\end{figure}

\begin{table}[!htb]
\centering
\begin{tabular}{ll|ccc}
  \hline
 && $\sigma_{\beta}^2=1$  & $\sigma_{\beta}^2=20$  & $\sigma_{\beta}^2=100$ \\ 
  \hline
  \rowcolor{lightGray}
 \cellcolor{white}
scenario 1 & mean & 0.997 & 1.000 & 1.000 \\
\rowcolor{Gray2}
 \cellcolor{white} & sd & 0.003 & 0.002 & 0.001 \\
 \rowcolor{lightGray}
 \cellcolor{white}
scenario 2 & mean & 0.999 & 0.991 & 0.995 \\ 
\rowcolor{Gray2}
 \cellcolor{white} & sd & 0.002 & 0.039 & 0.030 \\ 
 \rowcolor{lightGray}
 \cellcolor{white}
scenario 3 & mean & 0.994 & 1.000 & 0.998 \\ 
\rowcolor{Gray2}
 \cellcolor{white} & sd & 0.005 & 0.001 & 0.020 \\ 
 \rowcolor{lightGray}
 \cellcolor{white}
  scenario 4  & mean & 0.970 & 0.993 & 0.993 \\ 
  \rowcolor{Gray2}
  \cellcolor{white}& sd & 0.046 & 0.032 & 0.032 \\ 
  \rowcolor{lightGray}
 \cellcolor{white}
scenario 5 & mean & 0.986 & 0.990 & 0.976 \\ 
\rowcolor{Gray2}
 \cellcolor{white} & sd & 0.009 & 0.020 & 0.035 \\ 
 \rowcolor{lightGray}
 \cellcolor{white}
scenario 6 & mean & 0.898 & 0.976 & 0.971 \\ 
\rowcolor{Gray2}
 \cellcolor{white} & sd & 0.060 & 0.042 & 0.059 \\ 
 \rowcolor{lightGray}
 \cellcolor{white}
scenario 7 & mean & 1.000 & 0.980 & 0.929 \\ 
\rowcolor{Gray2}
 \cellcolor{white} & sd & 0.000 & 0.102 & 0.203 \\ 
   \hline
\end{tabular}
\caption{Mean and empirical standard deviation (sd) of the adjusted Rand index for the different settings of the sensitivity analysis.}
\label{table:ARI_SA_r1}
\end{table}

\section{Run time comparison}
\label{sec:app_runtime}
We compared the run time for all simulated seven settings across different sample sizes (100,250,500,1000). Table~\ref{table:run_time} below reports the comparison of the run time for BART, BCF, and CDBMM. BART has a superior performance in each scenario and sample size, while BCF and CDBMM have similar run times for the smaller sample size while the time difference increases with the sample size. However, the run times reported are measured in seconds for 1,000 iterations, which means that the complex time to run the models, including CDBMM, is not time-demanding and all the models have the same time magnitude.  Moreover, we want to underline that the times for the CDBMM include also the time to compute the point estimate of the partition, which uses the \texttt{R} package \texttt{mcclust} that requires a few seconds that increase with the sample size, while for BART and BCF we have not included the time to compute the CART. The current Gibbs sampler used to estimate the CDBMM is in plain R language while BCF is mainly written in C++ making the relative difference more a matter of software engineering than statistical modeling. 

\begin{table}[ht]
\centering
\footnotesize
\begin{tabular}{ll|ccccccc}
\hline
  && Scen. 1 & Scen. 2 & Scen. 3 & Scen. 4 & Scen. 5 & Scen. 6 & Scen. 7 \\
\hline
\rowcolor{lightGray}
 \cellcolor{white} 100 & BART & 2.18 & 2.21 & 2.07 & 2.54 & 2.43 & 2.88 & 2.21 \\ 
\rowcolor{Gray2}
 \cellcolor{white}  & BCF & 31.04 & 31.81 & 28.89 & 31.97 & 31.28 & 33.42 & 31.18 \\ 
\rowcolor{lightGray}
 \cellcolor{white}  & CDBMM & 31.93 & 30.45 & 31.22 & 32.11 & 34.80 & 39.42 & 30.02 \\ 
\rowcolor{Gray2}
\hline
 \cellcolor{white} 250 & BART & 3.43 & 4.47 & 3.56 & 3.93 & 3.74 & 3.47 & 3.90 \\ 
\rowcolor{lightGray}
 \cellcolor{white}  & BCF & 34.01 & 41.82 & 38.64 & 36.24 & 34.59 & 35.43 & 35.77 \\ 
\rowcolor{Gray2}
 \cellcolor{white}  & CDBMM & 68.66 & 77.29 & 71.04 & 74.70 & 73.60 & 72.07 & 72.84 \\ 
   \hline
 \rowcolor{lightGray}
 \hline
 \cellcolor{white} 500 & BART & 5.45 & 5.52 & 5.48 & 5.62 & 6.27 & 6.29 & 6.11 \\ 
\rowcolor{Gray2}
 \cellcolor{white}  & BCF & 39.43 & 38.82 & 42.86 & 38.20 & 40.40 & 41.21 & 46.08 \\
\rowcolor{lightGray}
 \cellcolor{white}  & CDBMM & 139.15 & 137.45 & 129.81 & 136.17 & 151.59 & 151.28 & 136.19 \\
 \rowcolor{Gray2}
 \hline
 \cellcolor{white} 1000 & BART & 11.27 & 10.82 & 10.35 & 11.76 & 11.35 & 12.72 & 9.67 \\  
\rowcolor{lightGray}
 \cellcolor{white}  & BCF & 50.75 & 52.89 & 48.43 & 55.76 & 51.08 & 51.71 & 50.40 \\
\rowcolor{Gray2}
 \cellcolor{white}  & CDBMM & 312.65 & 319.39 & 282.94 & 281.93 & 630.60 & 276.10 & 258.64 \\ 
\hline
\end{tabular}
\caption{Comparison of run time across different sample sizes (100,250,500,1000) and the 7 scenarios, for BART, BCF, and CDBMM. Time in seconds for 1000 interactions for one sample.}
\label{table:run_time}
\end{table}

\section{Dichotomization of exposure to \PMns}

On January 27, 2023, the EPA announced a proposal to lower the annual to reduce the annual PM2.5 National Ambient Air Quality Standard (NAAQS) in a range of 9 to 10 $\mu g/m^3$ \citep{epa2022}. To provide relevant information to current EPA regulatory decision-making we started exploring the potential harm associated with exposure to PM$_{2.5}$ levels surpassing the new proposed threshold for NAAQS. Thus, in our application, we dichotomize the exposure to PM$_{2.5}$ above and below 10 $\mu g/m^3$. In this specific context, our central objective differs from the estimation of the causal effect of continuous exposure on the outcome, which has been extensively investigated in prior studies \cite[see, e.g.,][]{josey2023air}. Our approach underscores that, although the exposure variable inherently maintains its continuous nature, we treat it as binary to effectively address this critical policy question.

In the context of heterogeneous treatment effect discovery, dichotomizing a continuous exposure could be a concern in situations where the dichotomization might conceal extremely larger/smaller exposure levels that could, in turn, result in larger/smaller heterogeneous treatment effects. We believe that a way to assess whether these concerns are potentially present in our data is twofold. 

First, one could examine the dispersion of the underlying PM$_{2.5}$ levels, and pay particular attention to the possibility of these levels taking exceptionally high or low values, which could raise concerns, especially if such values remain extreme even after the matching design step. To assess the distribution of PM$_{2.5}$ exposure in our Texas sample, we conducted a comparative analysis with the national PM$_{2.5}$ distribution to investigate any signs of overdispersion. Our empirical examination revealed that PM$_{2.5}$ levels in Texas, based on our dataset, exhibited a range spanning from 4.3$\mu g/m^3$ to 14.3$\mu g/m^3$. In contrast, the national PM$_{2.5}$ range for the same years extends from 1.4$\mu g/m^3$ to 18.1$\mu g/m^3$. Consequently, the PM$_{2.5}$ range within our Texas sample appears notably narrower than the broader national distribution, which mitigates the likelihood of extreme effects stemming from exceptionally high exposure levels. Furthermore, the post-matching distribution of underlying PM$_{2.5}$ exposure in our analysis is even narrower. Specifically, approximately 70\% of the analyzed zip codes exhibit PM$_{2.5}$ values within the range of $[8.5, 11.5]$,  90\% fall within the range of $[7, 12]$, and 98\% within the range of $[6, 13]$. Finally, our assessment indicates that the distribution of PM$_{2.5}$ exposure does not exhibit extreme values following the matching process. 

Second, one could check whether the larger/smaller estimated treatment effects correspond to units with larger/smaller values of the underlying PM$_{2.5}$ exposure. If one finds evidence that larger/smaller treatment effects are associated with larger/smaller PM$_{2.5}$ exposure values, this may be indicative of the dichotomization masking heterogeneous treatment effects due to extreme exposure levels. In our analysis of the estimated causal effect at the unit level using CDBMM, we find that the extreme values of PM$_{2.5}$ exposure align with values of treatment effects that fall within the center of their distribution. Conversely, when examining the extreme values of the causal effect at the unit level, we observe that they tend to be associated with units where PM$_{2.5}$ levels exhibit values towards the center of the exposure distribution. Thus, we find no empirical evidence of larger/smaller estimated treatment effects corresponding to units with larger/smaller values of the underlying PM$_{2.5}$ exposure.

\section{Study Design}
Our proposed model is applied to a matched dataset, where the census and meteorological variables are used for the matching. Matching is commonly used in observational studies to adjust for potential measured confounding bias \citep{rosenbaum1983central}. In this context, similarly to what has already been done in the literature on air pollution effects on health \cite[see, e.g., ][]{lee2021discovering, wu2020evaluating}, we decide to use matching before running our model to make our analyses as robust as possible with respect to potential measured confounding bias. 

We employ a 1-to-1 nearest neighbor propensity score matching, obtaining 1,402 selected units. The reduction of units is due to the different sample sizes of the treated and control groups in the original data, and 1-to-1 matching creates a sample with the same size for the treated and control groups. Using matching greatly improves the covariates balance. In Figure 3, we depict the covariate balance before and after the matching. Before the matching, there was a significant difference between the difference in standardized means of the observed values of some covariates, like poverty, education, or median household income, for the treated and control groups. These imbalances in the data might have led to spurious discoveries of effect heterogeneity. After the matching, the mean standardized differences of the covariates and the propensity score are included in the interval $[-0.1,0.1]$, which is usually used as a rule-of-thumb for good quality matches \citep{ho2007matching, austin2011introduction}. 

\begin{figure}
\begin{center}
\includegraphics[width=3in]{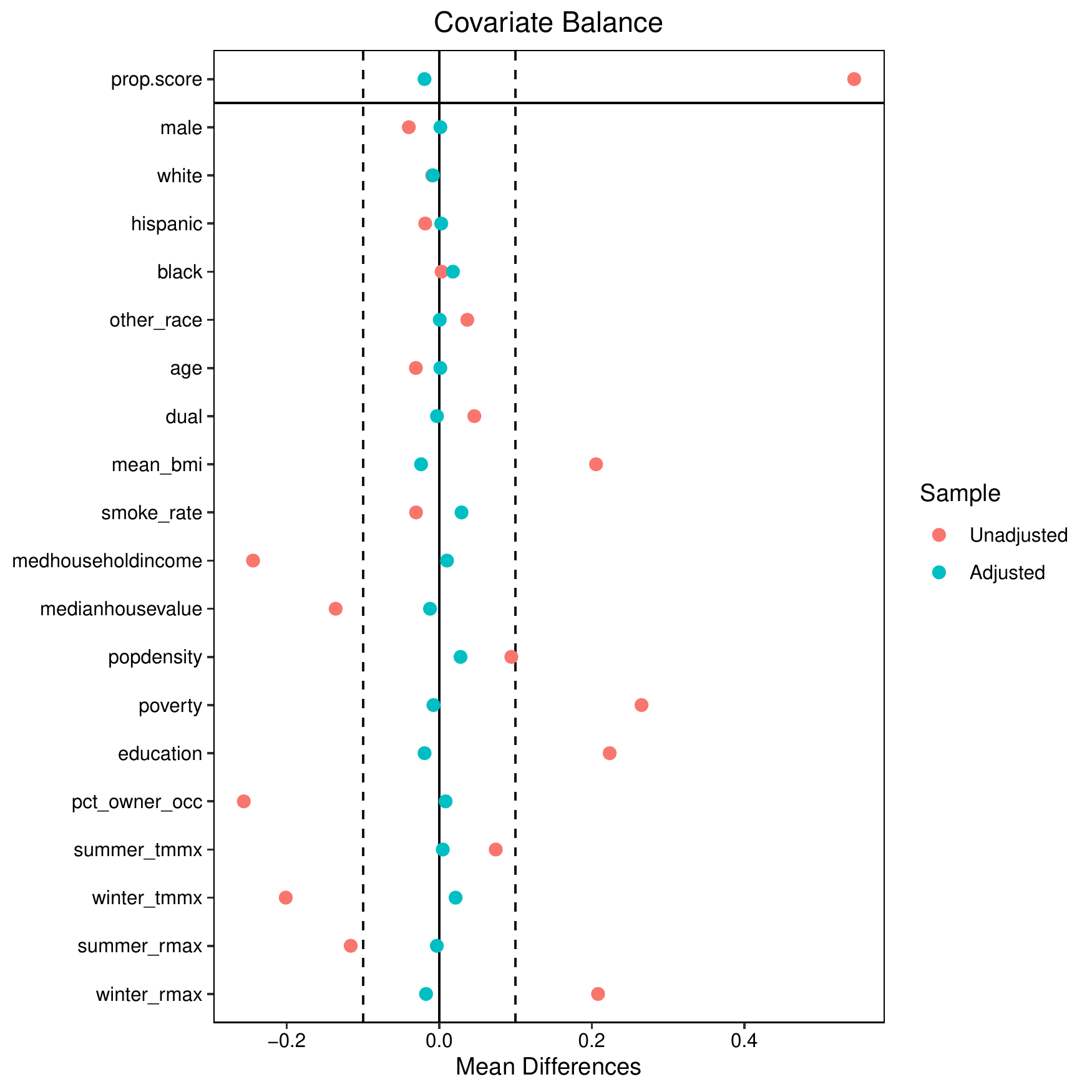}
\end{center}
\caption{Comparison of the covariate balance between before and after the nearest neighbor propensity score matching 1-to-1. The continuous black vertical line indicates the value 0, while the two dotted lines are for the values $-0.1$ and $0.1$, respectively.\label{fig:loveplot}}
\end{figure}

\section{Additional results for the Application: group average treatment effect}

As introduced in Section 2.3, our proposed method allows us to define and estimate any functions of the potential outcomes conditional to the group allocation. In particular, in the application, reported in Section 4, we also are interested in the Group Average Risk Ratio (GARR). This causal estimand is defined for the group $g$ as:
\begin{equation}
    GARR_g = \frac{\E \left[Y_i (1) \mid G_i= g \right]}{\E \left[Y_i(0) \mid G_i = g \right]},
    \label{eq_GARR}
\end{equation}
where $G_i$ is the Cartesian product of the point estimate of the cluster partition of $S_i^{(0)}$ and $S_i^{(1)}$, following the definition of the GATE in Section 2.3. Assuming the SUTVA and strong ignorability, GARR can be expressed as
\begin{equation}
    GARR_g = \frac{\E \left[Y_i \mid G_i= g, T_i=1 \right]}{\E \left[Y_i \mid G_i = g, T_i=0 \right]}.
\end{equation}

Basically, in our application, GARR defines the ratio between the mean of the posterior distribution of the outcomes under exposure to high levels of air pollution and the mean of the posterior distribution of the outcomes under lower levels of air pollution, for the units allocated in each identified group. This estimator helps us to describe the results in terms of percentage increment/decrement of the mortality rate when the units of the group are exposed to high level of \PM instead of low level of \PMns.

\begin{figure}[h]
\begin{center}
\includegraphics[width=4.5in]{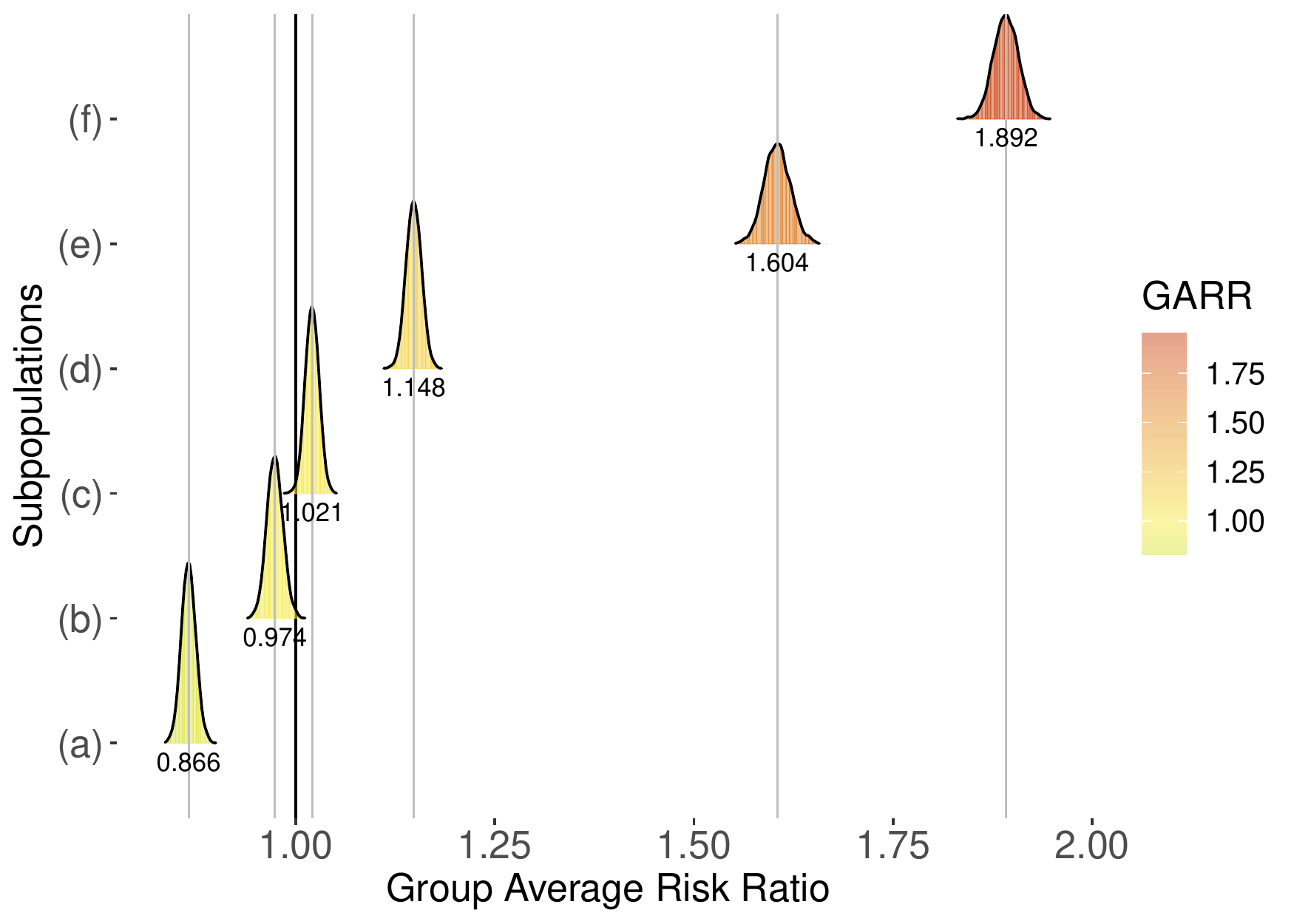}
\end{center}
\caption{Posterior distribution of GARR for the six estimated groups in the ZIP codes. The black line identifies the null causal effects and the gray lines are the mean of each posterior distribution for the GARRs.  \label{fig:GARR}}
\end{figure}

Figure \ref{fig:GARR} reports the distribution of the GARRs for the six identified groups, where the value 1 indicates the null causal effect of the exposure to \PM in the mortality rate.

The four groups, where exposure to higher levels of \PM increases the mortality rate, have values of GARR bigger than 1. Specifically, in the bigger group (c) the mortality rate of the population increases by $2\%$ under a high level of \PMns, followed by the group (d) with an increment of $15\%$, and the smaller groups (e) and (f) have an increment of $60\%$ and $89\%$ of the mortality rate, respectively, when exposed to a higher level of pollution. The two groups with an opposite trend show that the mortality rate of the population in these groups decreases by $2\%$ and $13\%$, respectively, under a high level of \PM instead of a lower level.

The results of GATE (reported in Section 4.3) and GARR  provide complementary information and furnish a deeper insight into the heterogeneity in the causal effects in the case of our application.

\section{Additional results for the Application: spacial distribution of groups}
\label{sec:app_map}

Here we look at the spatial distribution of the groups identified via our CDBMM model. In particular, we find that the groups characterized by higher vulnerability---i.e. groups (c), (d), (e), and (f)---are mostly located in southern Texas. The higher-vulnerability clusters are also found in suburban areas and along interstate highways between cities. Conversely, resilient groups---i.e. (a) and (b)---can be found in more sparsely populated areas. Gray areas could not be matched and thus are excluded from our analysis.

\begin{figure}[H]
\begin{center}
\includegraphics[trim={2cm 11.3cm 2cm 3.5cm}, width=6in]{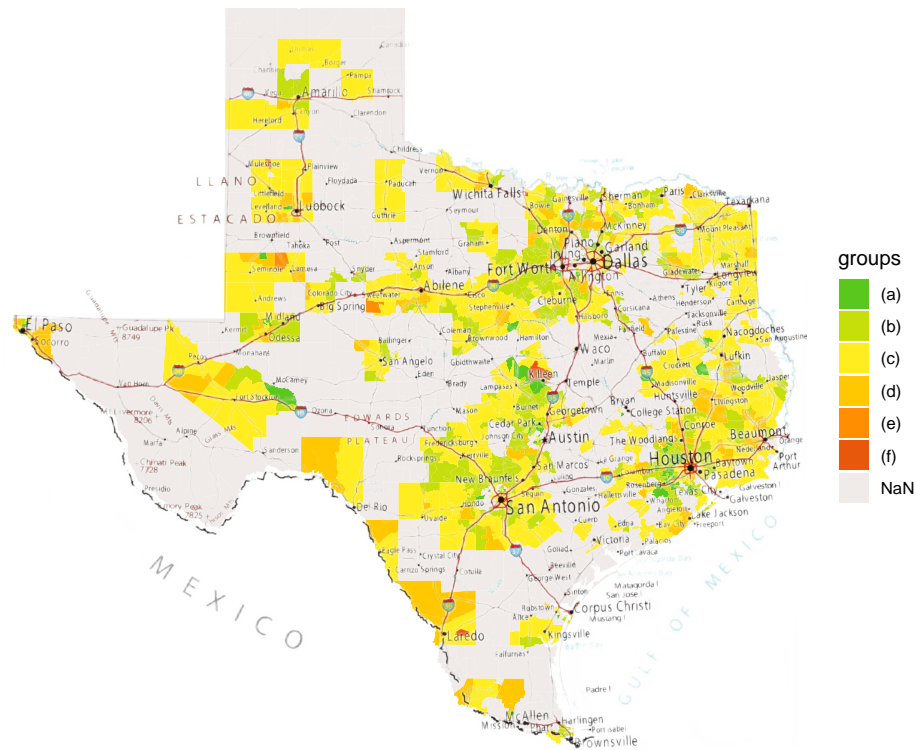}
\end{center}
\caption{Representation of the identified clusters on the map of Texas. \label{fig:map_cluster}}
\end{figure}

%% file: JASA-template.bbl
\begin{thebibliography}{}

\bibitem[\protect\citeauthoryear{Albert and Chib}{Albert and
  Chib}{2001}]{albert2001sequential}
Albert, J.~H. and S.~Chib (2001).
\newblock Sequential ordinal modeling with applications to survival data.
\newblock {\em Biometrics\/}~{\em 57\/}(3), 829--836.

\bibitem[\protect\citeauthoryear{Austin}{Austin}{2011}]{austin2011introduction}
Austin, P.~C. (2011).
\newblock An introduction to propensity score methods for reducing the effects
  of confounding in observational studies.
\newblock {\em Multivariate Behavioral Research\/}~{\em 46\/}(3), 399--424.

\bibitem[\protect\citeauthoryear{Bargagli-Stoffi, De-Witte, and
  Gnecco}{Bargagli-Stoffi et~al.}{2022}]{bargagli2022heterogeneous}
Bargagli-Stoffi, F.~J., K.~De-Witte, and G.~Gnecco (2022).
\newblock Heterogeneous causal effects with imperfect compliance: a {B}ayesian
  machine learning approach.
\newblock {\em The Annals of Applied Statistics\/}~{\em 16\/}(3), 1986--2009.

\bibitem[\protect\citeauthoryear{Bargagli~Stoffi, Garcia, Delaney, Deziel,
  Bell, and Dominici}{Bargagli~Stoffi et~al.}{2023}]{bargagli2023}
Bargagli~Stoffi, F.~J., D.~Garcia, S.~Delaney, N.~Deziel, M.~Bell, and
  F.~Dominici (2023).
\newblock Who is most vulnerable? causal machine learning to assess the
  heterogeneous health impacts of extreme heat exposure among elderly
  populations in north carolina and michigan.
\newblock {\em Working Paper\/}.

\bibitem[\protect\citeauthoryear{Binder}{Binder}{1978}]{binder1978bayesian}
Binder, D.~A. (1978).
\newblock Bayesian cluster analysis.
\newblock {\em Biometrika\/}~{\em 65\/}(1), 31--38.

\bibitem[\protect\citeauthoryear{Breiman, Friedman, Olshen, and Stone}{Breiman
  et~al.}{1984}]{breiman1984cart}
Breiman, L., J.~Friedman, R.~Olshen, and C.~Stone (1984).
\newblock Cart: Classification and regression trees.
\newblock {\em Wadsworth and Brooks/Cole Monterey, CA, USA\/}.

\bibitem[\protect\citeauthoryear{Carone, Dominici, and Sheppard}{Carone
  et~al.}{2020}]{carone2020pursuit}
Carone, M., F.~Dominici, and L.~Sheppard (2020).
\newblock In pursuit of evidence in air pollution epidemiology: the role of
  causally driven data science.
\newblock {\em Epidemiology (Cambridge, Mass.)\/}~{\em 31\/}(1), 1.

\bibitem[\protect\citeauthoryear{Chipman, George, and McCulloch}{Chipman
  et~al.}{2010}]{chipman2010bart}
Chipman, H.~A., E.~I. George, and R.~E. McCulloch (2010).
\newblock B{ART}: {B}ayesian additive regression trees.
\newblock {\em The Annals of Applied Statistics\/}~{\em 4\/}(1), 266--298.

\bibitem[\protect\citeauthoryear{Delaney, Mock, Mork, Cadei, Benmarhnia, Gill,
  Bell, Dominici, Bargagli~Stoffi, and Zanobetti}{Delaney
  et~al.}{2023}]{delaney2023}
Delaney, S., L.~Mock, D.~Mork, R.~Cadei, T.~Benmarhnia, T.~Gill, M.~Bell,
  D.~Dominici, Braun, F.~J. Bargagli~Stoffi, and A.~Zanobetti (2023).
\newblock Data-driven identification of risk factors for hospitalization with
  adrd from long-term exposure to fine particulate matter among medicare
  recipients.
\newblock {\em Working Paper\/}.

\bibitem[\protect\citeauthoryear{Dominici, Bargagli-Stoffi, and
  Mealli}{Dominici et~al.}{2021}]{dominici2020controlled}
Dominici, F., F.~J. Bargagli-Stoffi, and F.~Mealli (2021).
\newblock From controlled to undisciplined data: estimating causal effects in
  the era of data science using a potential outcome framework.
\newblock {\em Harvard Data Science Review\/}.

\bibitem[\protect\citeauthoryear{Dorie, Hill, Shalit, Scott, and Cervone}{Dorie
  et~al.}{2019}]{dorie2019automated}
Dorie, V., J.~Hill, U.~Shalit, M.~Scott, and D.~Cervone (2019).
\newblock Automated versus do-it-yourself methods for causal inference: Lessons
  learned from a data analysis competition.
\newblock {\em Statistical Science\/}~{\em 34\/}(1), 43--68.

\bibitem[\protect\citeauthoryear{Escobar and West}{Escobar and
  West}{1995}]{escobar1995bayesian}
Escobar, M.~D. and M.~West (1995).
\newblock Bayesian density estimation and inference using mixtures.
\newblock {\em Journal of the American Statistical Association\/}~{\em
  90\/}(430), 577--588.

\bibitem[\protect\citeauthoryear{Hahn, Murray, and Carvalho}{Hahn
  et~al.}{2020}]{hahn2020bayesian}
Hahn, P.~R., J.~S. Murray, and C.~M. Carvalho (2020).
\newblock Bayesian regression tree models for causal inference:
  {R}egularization, confounding, and heterogeneous effects (with discussion).
\newblock {\em Bayesian Analysis\/}~{\em 15\/}(3), 965--1056.

\bibitem[\protect\citeauthoryear{Hern{\'a}ndez, Raftery, Pennington, and
  Parnell}{Hern{\'a}ndez et~al.}{2018}]{hernandez2018bayesian}
Hern{\'a}ndez, B., A.~E. Raftery, S.~R. Pennington, and A.~C. Parnell (2018).
\newblock Bayesian additive regression trees using {B}ayesian model averaging.
\newblock {\em Statistics and Computing\/}~{\em 28\/}(4), 869--890.

\bibitem[\protect\citeauthoryear{Hill}{Hill}{2011}]{hill2011bayesian}
Hill, J.~L. (2011).
\newblock Bayesian nonparametric modeling for causal inference.
\newblock {\em Journal of Computational and Graphical Statistics\/}~{\em
  20\/}(1), 217--240.

\bibitem[\protect\citeauthoryear{Ho, Imai, King, and Stuart}{Ho
  et~al.}{2007}]{ho2007matching}
Ho, D.~E., K.~Imai, G.~King, and E.~A. Stuart (2007).
\newblock Matching as nonparametric preprocessing for reducing model dependence
  in parametric causal inference.
\newblock {\em Political Analysis\/}~{\em 15\/}(3), 199--236.

\bibitem[\protect\citeauthoryear{Holland}{Holland}{1986}]{holland1986statistics}
Holland, P.~W. (1986).
\newblock Statistics and causal inference.
\newblock {\em Journal of the American Statistical Association\/}~{\em
  81\/}(396), 945--960.

\bibitem[\protect\citeauthoryear{Jacob}{Jacob}{2019}]{jacob2019group}
Jacob, D. (2019).
\newblock Group average treatment effects for observational studies.
\newblock {\em arXiv preprint arXiv:1911.02688\/}.

\bibitem[\protect\citeauthoryear{Jbaily, Zhou, Liu, Lee, Kamareddine, Verguet,
  and Dominici}{Jbaily et~al.}{2022}]{jbaily2022air}
Jbaily, A., X.~Zhou, J.~Liu, T.-H. Lee, L.~Kamareddine, S.~Verguet, and
  F.~Dominici (2022).
\newblock Air pollution exposure disparities across us population and income
  groups.
\newblock {\em Nature\/}~{\em 601\/}(7892), 228--233.

\bibitem[\protect\citeauthoryear{Josey, Delaney, Wu, Nethery, DeSouza, Braun,
  and Dominici}{Josey et~al.}{2023}]{josey2023air}
Josey, K.~P., S.~W. Delaney, X.~Wu, R.~C. Nethery, P.~DeSouza, D.~Braun, and
  F.~Dominici (2023).
\newblock Air pollution and mortality at the intersection of race and social
  class.
\newblock {\em New England Journal of Medicine\/}~{\em 388\/}(15), 1396--1404.

\bibitem[\protect\citeauthoryear{Krantsevich, He, and Hahn}{Krantsevich
  et~al.}{2023}]{krantsevich2023stochastic}
Krantsevich, N., J.~He, and P.~R. Hahn (2023).
\newblock Stochastic tree ensembles for estimating heterogeneous effects.
\newblock In {\em International Conference on Artificial Intelligence and
  Statistics}, pp.\  6120--6131. PMLR.

\bibitem[\protect\citeauthoryear{Lee, Bargagli-Stoffi, and Dominici}{Lee
  et~al.}{2020}]{lee2020causal}
Lee, K., F.~J. Bargagli-Stoffi, and F.~Dominici (2020).
\newblock Causal rule ensemble: Interpretable inference of heterogeneous
  treatment effects.
\newblock {\em arXiv preprint arXiv:2009.09036\/}.

\bibitem[\protect\citeauthoryear{Lee, Small, and Dominici}{Lee
  et~al.}{2021}]{lee2021discovering}
Lee, K., D.~S. Small, and F.~Dominici (2021).
\newblock Discovering heterogeneous exposure effects using randomization
  inference in air pollution studies.
\newblock {\em Journal of the American Statistical Association\/}~{\em
  116\/}(534), 569--580.

\bibitem[\protect\citeauthoryear{Li, Konisky, and Zirogiannis}{Li
  et~al.}{2019}]{li2019racial}
Li, Z., D.~M. Konisky, and N.~Zirogiannis (2019).
\newblock {Racial, ethnic, and income disparities in air pollution: A study of
  excess emissions in Texas}.
\newblock {\em PloS One\/}~{\em 14\/}(8), e0220696.

\bibitem[\protect\citeauthoryear{Linero and Antonelli}{Linero and
  Antonelli}{2023}]{linero2021and}
Linero, A.~R. and J.~L. Antonelli (2023).
\newblock The how and why of {B}ayesian nonparametric causal inference.
\newblock {\em Wiley Interdisciplinary Reviews: Computational
  Statistics\/}~{\em 15\/}(1), e1583.

\bibitem[\protect\citeauthoryear{Linero and Yang}{Linero and
  Yang}{2018}]{linero2018bayesian1}
Linero, A.~R. and Y.~Yang (2018).
\newblock Bayesian regression tree ensembles that adapt to smoothness and
  sparsity.
\newblock {\em Journal of the Royal Statistical Society: Series B (Statistical
  Methodology)\/}~{\em 80\/}(5), 1087--1110.

\bibitem[\protect\citeauthoryear{MacEachern}{MacEachern}{2000}]{mac2000dependent}
MacEachern, S.~N. (2000).
\newblock Dependent {D}irichlet processes.
\newblock {\em Technical Report. Department of Statistics, The Ohio State
  University, Columbus, OH.\/}.

\bibitem[\protect\citeauthoryear{Meil{\u{a}}}{Meil{\u{a}}}{2007}]{meilua2007comparing}
Meil{\u{a}}, M. (2007).
\newblock Comparing clusterings—an information based distance.
\newblock {\em Journal of Multivariate Analysis\/}~{\em 98\/}(5), 873--895.

\bibitem[\protect\citeauthoryear{Oganisian, Mitra, and Roy}{Oganisian
  et~al.}{2021}]{oganisian2021bayesian}
Oganisian, A., N.~Mitra, and J.~A. Roy (2021).
\newblock A {B}ayesian nonparametric model for zero-inflated outcomes:
  Prediction, clustering, and causal estimation.
\newblock {\em Biometrics\/}~{\em 77\/}(1), 125--135.

\bibitem[\protect\citeauthoryear{Quintana}{Quintana}{2006}]{quintana2006predictive}
Quintana, F.~A. (2006).
\newblock A predictive view of {B}ayesian clustering.
\newblock {\em Journal of Statistical Planning and Inference\/}~{\em 136\/}(8),
  2407--2429.

\bibitem[\protect\citeauthoryear{Quintana, M{\"u}ller, Jara, and
  MacEachern}{Quintana et~al.}{2022}]{quintana2020dependent}
Quintana, F.~A., P.~M{\"u}ller, A.~Jara, and S.~N. MacEachern (2022).
\newblock {The dependent Dirichlet process and related models}.
\newblock {\em Statistical Science\/}~{\em 37\/}(1), 24--41.

\bibitem[\protect\citeauthoryear{Rodriguez and Dunson}{Rodriguez and
  Dunson}{2011}]{rodriguez2011nonparametric}
Rodriguez, A. and D.~B. Dunson (2011).
\newblock Nonparametric {B}ayesian models through probit stick-breaking
  processes.
\newblock {\em Bayesian Analysis\/}~{\em 6}, 1.

\bibitem[\protect\citeauthoryear{Rosenbaum and Rubin}{Rosenbaum and
  Rubin}{1983}]{rosenbaum1983central}
Rosenbaum, P.~R. and D.~B. Rubin (1983).
\newblock The central role of the propensity score in observational studies for
  causal effects.
\newblock {\em Biometrika\/}~{\em 70\/}(1), 41--55.

\bibitem[\protect\citeauthoryear{Roy, Lum, Zeldow, Dworkin, Re~III, and
  Daniels}{Roy et~al.}{2018}]{roy2018bayesian}
Roy, J., K.~J. Lum, B.~Zeldow, J.~D. Dworkin, V.~L. Re~III, and M.~J. Daniels
  (2018).
\newblock Bayesian nonparametric generative models for causal inference with
  missing at random covariates.
\newblock {\em Biometrics\/}~{\em 74\/}(4), 1193--1202.

\bibitem[\protect\citeauthoryear{Rubin}{Rubin}{1974}]{rubin1974estimating}
Rubin, D.~B. (1974).
\newblock Estimating causal effects of treatments in randomized and
  nonrandomized studies.
\newblock {\em Journal of Educational Psychology\/}~{\em 66\/}(5), 688.

\bibitem[\protect\citeauthoryear{Rubin}{Rubin}{1980}]{rubin1980randomization}
Rubin, D.~B. (1980).
\newblock Randomization analysis of experimental data: The fisher randomization
  test comment.
\newblock {\em Journal of the American Statistical Association\/}~{\em
  75\/}(371), 591--593.

\bibitem[\protect\citeauthoryear{Rubin}{Rubin}{1986}]{rubin1986comment}
Rubin, D.~B. (1986).
\newblock Comment: Which ifs have causal answers.
\newblock {\em Journal of the American Statistical Association\/}~{\em
  81\/}(396), 961--962.

\bibitem[\protect\citeauthoryear{Sethuraman}{Sethuraman}{1994}]{sethuraman1994constructive}
Sethuraman, J. (1994).
\newblock A constructive definition of {D}irichlet priors.
\newblock {\em Statistica Sinica\/}~{\em 4}, 639--650.

\bibitem[\protect\citeauthoryear{Shaw, Hayes-Larson, Glymour, Dufouil, Hohman,
  Whitmer, Kobayashi, Brookmeyer, and Mayeda}{Shaw
  et~al.}{2021}]{shaw2021evaluation}
Shaw, C., E.~Hayes-Larson, M.~M. Glymour, C.~Dufouil, T.~J. Hohman, R.~A.
  Whitmer, L.~C. Kobayashi, R.~Brookmeyer, and E.~R. Mayeda (2021).
\newblock Evaluation of selective survival and sex/gender differences in
  dementia incidence using a simulation model.
\newblock {\em JAMA Network Open\/}~{\em 4\/}(3), e211001--e211001.

\bibitem[\protect\citeauthoryear{Sivaganesan, M{\"u}ller, and
  Huang}{Sivaganesan et~al.}{2017}]{sivaganesan2017subgroup}
Sivaganesan, S., P.~M{\"u}ller, and B.~Huang (2017).
\newblock Subgroup finding via {B}ayesian additive regression trees.
\newblock {\em Statistics in Medicine\/}~{\em 36\/}(15), 2391--2403.

\bibitem[\protect\citeauthoryear{U.S.\;Census\;Bureau}{U.S.\;Census\;Bureau}{2020}]{UScensus}
U.S.\;Census\;Bureau (2020).
\newblock {QuickFacts: Texas}.
\newblock \url{https://www.census.gov/quickfacts/TX}.

\bibitem[\protect\citeauthoryear{U.S.\;Environmental\;Protection\;Agency}{U.S.\;Environmental\;Protection\;Agency}{2022}]{epa2022}
U.S.\;Environmental\;Protection\;Agency (2022).
\newblock Regulatory impact analysis for the proposed reconsideration of the
  national ambient air quality standards for particulate matter.
\newblock {\em Technical Report: EPA-452/P-22-001\/}.

\bibitem[\protect\citeauthoryear{Wade, Dunson, Petrone, and Trippa}{Wade
  et~al.}{2014}]{wade2014improving}
Wade, S., D.~B. Dunson, S.~Petrone, and L.~Trippa (2014).
\newblock Improving prediction from {D}irichlet process mixtures via
  enrichment.
\newblock {\em The Journal of Machine Learning Research\/}~{\em 15\/}(1),
  1041--1071.

\bibitem[\protect\citeauthoryear{Wade and Ghahramani}{Wade and
  Ghahramani}{2018}]{wade2018bayesian}
Wade, S. and Z.~Ghahramani (2018).
\newblock Bayesian cluster analysis: {P}oint estimation and credible balls.
\newblock {\em Bayesian Analysis\/}~{\em 13\/}(2), 559--626.

\bibitem[\protect\citeauthoryear{Wendling, Jung, Callahan, Schuler, Shah, and
  Gallego}{Wendling et~al.}{2018}]{wendling2018comparing}
Wendling, T., K.~Jung, A.~Callahan, A.~Schuler, N.~H. Shah, and B.~Gallego
  (2018).
\newblock Comparing methods for estimation of heterogeneous treatment effects
  using observational data from health care databases.
\newblock {\em Statistics in Medicine\/}~{\em 37\/}(23), 3309--3324.

\bibitem[\protect\citeauthoryear{Wu, Braun, Schwartz, Kioumourtzoglou, and
  Dominici}{Wu et~al.}{2020}]{wu2020evaluating}
Wu, X., D.~Braun, J.~Schwartz, M.~Kioumourtzoglou, and F.~Dominici (2020).
\newblock Evaluating the impact of long-term exposure to fine particulate
  matter on mortality among the elderly.
\newblock {\em Science Advances\/}~{\em 6\/}(29), eaba5692.

\bibitem[\protect\citeauthoryear{Yeager, Hanselman, Walton, Murray, Crosnoe,
  Muller, Tipton, Schneider, Hulleman, Hinojosa, et~al.}{Yeager
  et~al.}{2019}]{yeager2019national}
Yeager, D.~S., P.~Hanselman, G.~M. Walton, J.~S. Murray, R.~Crosnoe, C.~Muller,
  E.~Tipton, B.~Schneider, C.~S. Hulleman, C.~P. Hinojosa, et~al. (2019).
\newblock A national experiment reveals where a growth mindset improves
  achievement.
\newblock {\em Nature\/}~{\em 573\/}(7774), 364--369.

\end{thebibliography}
